\documentclass[
twocolumn,
aps,superscriptaddress,pra]{revtex4-1}
\usepackage{color}
\usepackage{textcomp}
\usepackage{amsmath}
\usepackage{amssymb}
\usepackage{graphicx}
\usepackage{esint}
\usepackage{amsfonts}
\usepackage{bm}
\usepackage{multirow}
\usepackage{natbib}
\usepackage[normalem]{ulem}
\usepackage{verbatim}
\usepackage{braket}
\usepackage{xr-hyper}
\usepackage{hyperref}
\usepackage{relsize}
\usepackage{comment}
\usepackage[version=3]{mhchem} 

\usepackage{times}
\hypersetup{colorlinks=true}
\usepackage[all]{hypcap} 

\makeatletter
\@ifundefined{textcolor}{}
{%
 \definecolor{BLACK}{gray}{0}
 \definecolor{WHITE}{gray}{1}
 \definecolor{RED}{rgb}{1,0,0}
 \definecolor{GREEN}{rgb}{0,1,0}
 \definecolor{BLUE}{rgb}{0,0,1}
 \definecolor{CYAN}{cmyk}{1,0,0,0}
 \definecolor{MAGENTA}{cmyk}{0,1,0,0}
 \definecolor{YELLOW}{cmyk}{0,0,1,0}
}

\@ifundefined{showcaptionsetup}{}
\makeatother
\usepackage{babel}
\def \be {\begin{equation}}
\def \ee {\end{equation}}
\def \bea {\begin{align}}
\def \eea {\end{align}}
\def \p {\partial}
\def \BEA {\begin{eqnarray}}
\def \EEA {\end{eqnarray}}

\def \m {\mathbf }
\begin{document}

\title{Heating of inhomogeneous electron flow in the hydrodynamic regime}

\author{Gu Zhang}
\email{gu.zhang@kit.edu}
\affiliation{Institute for Quantum Materials and Technologies, Karlsruhe Institute of Technology, 76021 Karlsruhe, Germany}

\author{Valentin Kachorovskii}
\affiliation{Ioffe Institute, 194021 
St.~Petersburg, Russia}

\author{Konstantin Tikhonov}
\affiliation{Skolkovo Institute of Science and Technology, Moscow, 121205, Russia}
\affiliation{Condensed-matter Physics Laboratory, National Research University Higher School of Economics, 101000 Moscow, Russia}

\author{Igor Gornyi}
\affiliation{Institute for Quantum Materials and Technologies, Karlsruhe Institute of Technology, 76021 Karlsruhe, Germany}
\affiliation{Ioffe Institute, 194021 
St.~Petersburg, Russia}

\begin{abstract}

We study the electron temperature profiles for an inhomogeneous electron flow in the hydrodynamic regime. We assume that the inhomogeneity is due to a weakly non-uniform distribution of the momentum relaxation time within a spherically constricted area. We show that the temperature profile dramatically depends on the drive strength and the viscosity of the electron liquid. In the absence of viscosity, a Landauer-dipole-like temperature distribution, asymmetrically deformed along the current by the inelastic electron-phonon scattering, emerges around the inhomogeneity. We find that both the Landauer-dipole temperature profile and its asymmetry in the direction of the driving  electric field exist in all dimensionalities and are, therefore,  universal features of inhomogeneous hydrodynamic electron flow. We further demonstrate that the electron viscosity suppresses the thermal Landauer dipole  and leads to the appearance of a ``hot spot'' exactly at the center of the constriction. We also calculate the phonon temperature distribution, which can be directly measured in experiments on thermal nanoimaging.

\end{abstract}

\date{\today}
\maketitle

\section{Introduction} 
\label{sec:intro}

The study of low-dimensional electronic systems is a key direction in the condensed matter physics in the last few decades. This is dictated by the general trend in the reduction of the sizes of the electronic devices, and is supported by significant advances in modern technology.  The semiconducting heterostructures \cite{IhnBook} and graphene \cite{Novoselov05} are among these technological developments, through which the two-dimensional (2D) electron gas has been experimentally realized and employed for designing nanoelectronic devices.  

One of the most important properties of an electronic circuit is its ability to cool efficiently and operate under a sufficiently strong drive, when transport becomes substantially nonequilibrium.  Under the non-equilibrium conditions, effects related to overheating, dissipation, and thermalization become decisive for the device functioning. The variety of setups in which such phenomena define the physical properties of the system is quite wide, and surprises arise even when studying more conventional structures subject to the drive. The need of developing a comprehensive theory of heat transfer in nanosystems has become particularly evident in recent years. Indeed, miniaturization of electronic devices down to the nanoscale and the use of new materials with unique properties are expected to affect the thermal properties of nanostructures in a crucial way. 

As the nanoscales are reached, new effects come into play owing to the increased role of  disorder, electron-electron interactions, and their interplay. Further, it is now possible to  change smoothly  the dimensionality of the system. For example, through etched gates, a 2D electron gas can be divided into multiple areas that are connected by point contacts \cite{Goldhaber-GordonNature98,CronenwettScience98}. Properly  selected configuration of gates allows one to create a contact in the form of a quasi-one-dimensional constriction and control the number of channels responsible for current transfer through such a constriction. By using a selective doping, one can  engineer inhomogeneous low-dimensional gate-controllable  structures to probe charge and heat transport.

At the same time, the development of the SQUID-on-tip (tSOT) \cite{FinklerNanoLett10,Vasyukov} and the cryogenic quantum magnetometry \cite{VoolX2020} techniques enables precise measurement of the temperature profiles and electric current distributions in nanostructures. The former technique has already been successfully applied for the imaging of impurities  \cite{HalbertalNat16,HalbertalScience17} and the quantum Hall edges states of graphene samples \cite{Zeldov2019,UriNatPhys20,AharonX20}. These modern experimental techniques can be applied to the analysis of the influence
of various types of nanoscale inhomogeneities (intrinsic, geometrical, artificial) on heat balance in nanodevices.

Recently, a new direction in the physics of low-dimensional systems has emerged---electronic hydrodynamics,  
which was discussed for many decades \cite{gurzhi1,gurzhi2,gurzhi,DeJongPRB95}, but was scarcely studied  because of the lack of experimental realization at that time. Now this direction is booming thanks to the technological advances in the production of ultraclean ballistic systems, and a number of new  hydrodynamic regimes were theoretically predicted (for review, see, e.g., Refs. \cite{NarozhnyAnnalenderPhysik17,LucasFongJPhys18,NarozhnyAofPhys19,PoliniGeimPhysicsToday20}) and experimentally discovered \cite{BandurinScience16, CrossnoScience16,MollScience16,GhahariPRL16,KumarNatPhys17,BandurinNatureComm18,BraemPRB18,AlexandreNPJQM18,GoothNatCom18,BerdyuginScience19,GallagherScience19,SulpizioNatPhys19,EllaNatNano19,MarkNature20,RaichevPRB20,GusevSciRep20,VoolX2020,GenursX2020,KimNatCom20,GuptaPRL21,GusevPRB21,ZhangShurMichaelJounalAppPhys21,JaouiNatCom21}. In particular, with the use of modern nanoimaging techniques, it became possible to visualize hydrodynamic flows in 2D materials \cite{BraemPRB18,EllaNatNano19,SulpizioNatPhys19,MarkNature20,VoolX2020}.

Despite the experimental advances in the thermal detection, the studies of the hydrodynamic phenomena and transition regimes between hydrodynamics, in which the electron-electron scattering is the fastest, and the drift-diffusion regime, where scattering by disorder dominates, mostly focus on the charge transport.
The heat transport features, however, are comparatively less visited \cite{RokniLevinson1995,GurevichPRB97,GatiyatovJETP10} and overheating of the sufficiently small  electron devices---the issue of crucial importance for possible applications---in fact, remains very poorly understood. Importantly, as was pointed out more than twenty years ago \cite{RokniLevinson1995}, the local Joule heating approximation does not work for the description of small devices like point contacts. The heat generation there can be  governed by nonlocal dissipation processes. However, nonlocality is not the only specific feature of the heat dissipation at the nanoscales.

\begin{figure}[th!]
        \includegraphics[width=0.83\columnwidth]{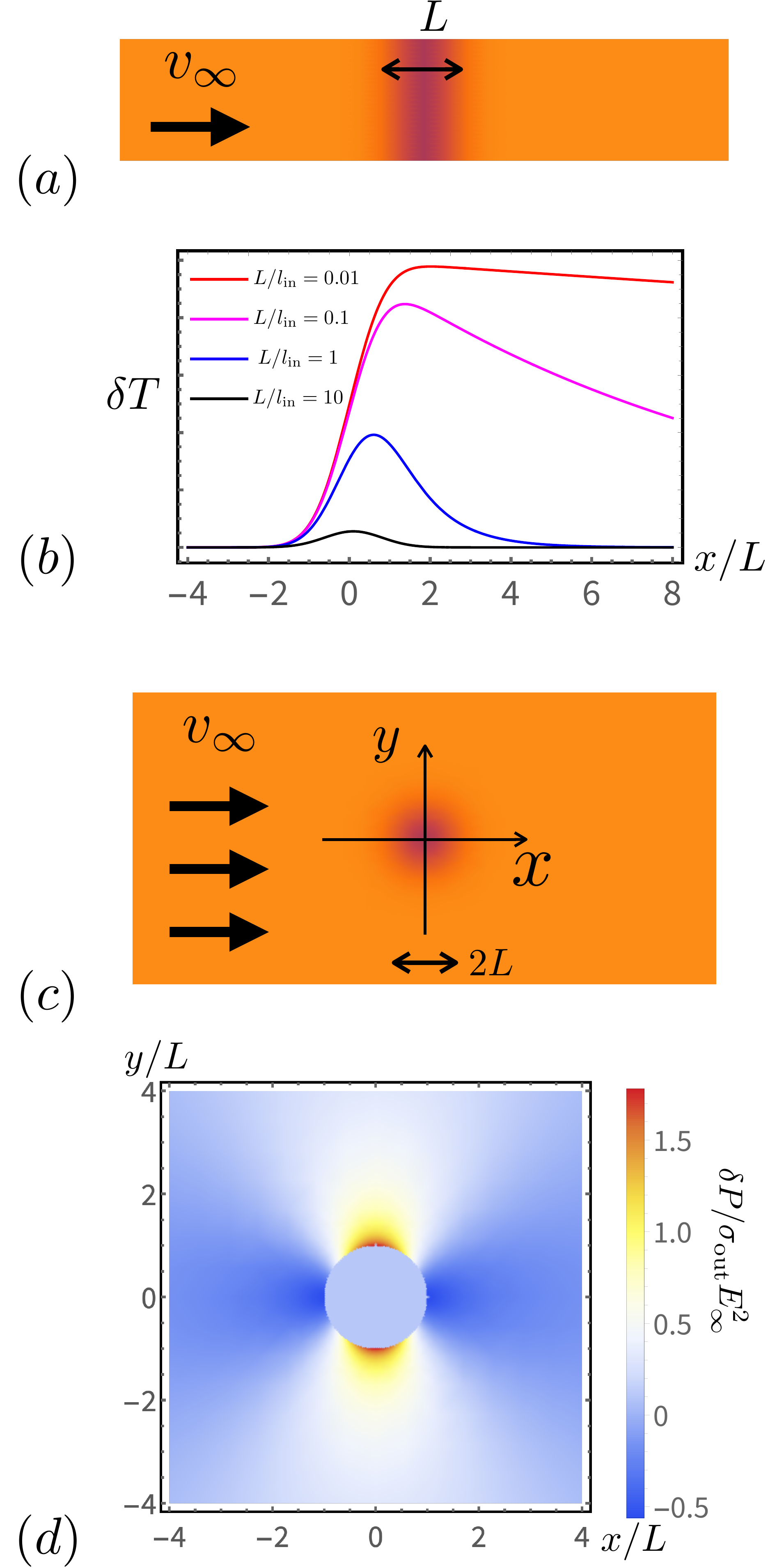}
  \caption{ (a) Quasi-1D symmetric  constriction with increased scattering rate and the corresponding  asymmetric  overheating profile (b). The temperature asymmetry increases when the ratio between the constriction size $L$ and the drift inelastic length $l_{\text{in}}$ decreases (see Ref.~\cite{Tikhonov2019}).  
The  circular-shape constriction with increased  impurity scattering in 2D or 3D electron gas (c) driven by the flow along the $x$ direction, with the homogeneous drift velocity $v_\infty$ at $r\to \infty$. The magnitude of the momentum relaxation time $\tau$ is shown  by the color (the lighter the color, the larger the value of $\tau$).
(d) Constriction-induced local heating in a 2D or 3D sample, calculated with the local-heating approximation, see Appendix \ref{app:multi_conduct}.}
      \label{fig:model}
\end{figure}

A recent analysis~\cite{Tikhonov2019} shows a great variety of different overheating regimes in a quasi-one-dimensional (quasi-1D) constriction [see Fig.~\ref{fig:model}(a)] with an inhomogeneous distribution of transport scattering rate. One of the distinguishing features of all these regimes  
is the presence of strongly asymmetric temperature profiles [see Fig.~\ref{fig:model}(b)]. The heat transfer in 2D graphene with local defects was addressed theoretically in Refs.~\cite{TikhononvPRB18,KongPRB18} within the concept of supercollisions---the impurity-stimulated electron-phonon scattering~\cite{SongReizerLevitovPRL12,SongLevitov} applied to resonant scatterers, as suggested by experimental observations \cite{HalbertalScience17}. The spatial distribution of dissipation power  was linked there to the formation of Landauer dipoles \cite{Landauer} around the local defect [cf. Fig.~\ref{fig:model}(d)]. However, only the total dissipation power was calculated in Ref.~\cite{TikhononvPRB18} (in the linear-response regime), without exploring the spatial structure of the local temperature profiles.

In this paper, we study in detail the thermal characteristics of an inhomogeneous 2D sample in the hydrodynamic regime, focusing on the temperature profiles induced by inhomogeneities. We also discuss the thermal properties of 3D  inhomogeneous systems within a simplified model.
The inhomogeneity is introduced by the presence of a constricted area, where the impurity scattering rate is different from its uniform value (i.e, its value far away from the constriction).
We consider a circular-shape constriction [Fig.~\ref{fig:model}(c)], where the constriction has a higher impurity scattering rate. Similarly to the quasi-1D case, we predict a strong asymmetry in the electronic temperature profiles,  even for a perfectly symmetric constriction. 

We assume  that the inhomogeneity is weak, and study the dc-current-induced  variation of temperature, density and drift velocity distributions.
One of our main results is the prediction of a the Landauer-dipole character of the temperature distribution (similar to the distribution of the electric field in the original Landauer's consideration~\cite{Landauer}). 
This Landauer-dipole temperature profile is further shifted asymmetrically with increasing the driving current, as in the quasi-1D setting~\cite{Tikhonov2019}.
Remarkably, this Landauer-dipole asymmetric feature universally exists in all dimensionalities
[see Fig.~\ref{fig:model}].
Our second key result is that the  temperature profile is dramatically modified by the viscosity of the electron liquid. We  demonstrate that the electron viscosity suppresses the Landauer-dipole-like  structure and creates a ``hot spot''  exactly at the center of the constriction.   
We also discuss  the corresponding phonon-temperature profiles that can be experimentally observed using the tSOT technique \cite{FinklerNanoLett10,HalbertalNat16}.

The paper is organized as follows.
In Sec.~\ref{sec:hydro_eqs}, we describe our model and present the basic  
equations.  In Sec.~\ref{sec:current}, we analyze the current-induced overheating for different dimensionalities, geometries, and interaction models within the ideal fluid model. In Sec.~\ref{viscosity}, we discuss the viscous case. Section~\ref{sec:phonon_temp} addresses the phonon temperature. Our results are summarized in Sec.~\ref{sec:summary}.  The details of calculations are discussed in Appendices~\ref{Gauss} and \ref{app:multi_conduct}.

\section{Model and basic equations}
\label{sec:hydro_eqs}

In this section, we introduce the hydrodynamic equations of ideal electron fluid (Sec.~\ref{sec:hydro}) and the constriction model (Sec.~\ref{sec:constriction}). The formalism adopted in the present paper closely follows and generalizes the one of Ref.~\cite{Tikhonov2019}, where an ideal-fluid description of a quasi-1D constriction was developed.

\subsection{Hydrodynamic approach}
\label{sec:hydro}
We consider a 2D spatially inhomogeneous, weakly disordered electron system, where the inhomogeneity is induced by the dependence of the electron transport scattering time $\tau$ on position $\m r$. 
The distribution function    $f(\mathbf r,  \mathbf V)$
obeys the stationary kinetic equation 
\begin{equation}
\m V   \frac{\partial f}{\partial \m r} + \frac{\m{F}}{m}  \frac{\partial f}{\partial \m{V}} = \widehat{\rm St}f.
\label{eq:kinetic_equation}
\end{equation}
Here $\m{V}$ and $m$ are the electron velocity and mass, respectively,   $\m {F}$  is the total local force.
The right-hand side (r.h.s.) of 
Eq.~(\ref{eq:kinetic_equation})
is determined by the collision  operator
$$\widehat{\rm St} = \widehat{\rm St}_{\text{\rm imp}} + \widehat{\rm St}_{\text{\rm ph}} + \widehat{\rm St}_{\text{\rm ee}},$$ 
which contains three terms describing electron-impurity, electron-phonon, and electron-electron scattering. 
The  impurity part of the collision integral reads 
\begin{equation}
\widehat{\rm St}_{\text{imp}} = \frac{f_0 - f}{\tau (\m{r})},
\label{eq:collision_integral}
\end{equation}
 where $f_0 = \langle f \rangle_\theta$ is the distribution function  averaged over the  velocity angle  $\theta$.
Below, we assume that the transport relaxation time $\tau (\m{r})$ is constant everywhere except for a certain constricted area. We model the electron-phonon  scattering by  the following Fokker-Planck-type collision integral  
\begin{equation}
\widehat{\rm St}_{\rm ph } = \gamma \frac{\partial}{\partial \epsilon} \left\{ \epsilon \left[ f_0 ( 1 - f_0 ) + T_0 \frac{\partial }{\partial \epsilon }f_0 \right]  \right\},
\label{eq:e_ph_intergral}
\end{equation}
where $\gamma$ is the electron-phonon scattering rate, $\epsilon$ is the single-particle energy, and $T_0$ is the phonon  temperature.  We assume $\gamma \tau \ll 1$. The specific form of the electron-electron collision operator,  $\widehat{\rm  St}_{\text{ee}}$, is not important to us. We consider the hydrodynamic limit where the electron-electron collision time, $\tau_{\text{ee}}$, is much shorter than the momentum relaxation time: $\tau_{\text{ee}} \ll \tau$.
We also assume that $\tau_{\text{ee}} \ll  1/\gamma.$

First, we  consider an ideal electron liquid neglecting the electron viscosity and the heat conductivity, which are both proportional to $\tau_{\rm ee}$. 
The viscous liquid will be discussed 
in Sec.~\ref{viscosity}.
In the limit $\tau_{\rm ee} \to 0$, one can use the hydrodynamic Ansatz for the distribution function 
\be
f(\m{r},\m{V}) = \frac{1}{\exp \left[ \frac{m [ \m{V} - \m{v}(\m{r}) ]^2/2 - \mu(\m{r}) }{T(\m{r })} \right] + 1},
\label{eq:hydro_ansatz}
\ee 
where  functions $\mu (\m{r})$, $T(\m{r})$, and $\m{v}(\m r)$ are the local values of  chemical potential, electron temperature, and the drift velocity, respectively.
The equations governing these collective (hydrodynamic) variables are obtained by averaging Eq.~\eqref{eq:kinetic_equation} multiplied by $1$, $\m{V}$, and $m V^2$, respectively, over the velocity $\m{V}$  (see, for example,  Appendix A in  Ref.~\cite{Tikhonov2019}).
After some algebra, we get the following set of equations:
\BEA
&&\!\!{\rm div}\left(\mathbf v N \right)=0,
\label{dN-dt}
\\
&&\!\!(\mathbf v \nabla) \mathbf v + \frac{\mathbf v}{\tau}=\frac{1}{m}\left( \mathbf{F}_0 + \delta \m F   - \frac{1}{N} \nabla W \right),
\label{dv-dt}
\\
&&
\!\! C~ {\rm div} (\mathbf v T)=N\left[\frac{mv^2}{\tau}-\gamma ~ (T-T_0) \right].
\label{dC-dt}
\EEA
Here, $N=N(\m r)$ is the local electron concentration,
related to the local chemical  potential  as follows ($\nu$ is the 2D density of states):
\be 
\mu=T\ln \left[ \exp\left (\frac{N}{ \nu T}\right)-1\right]. 
\label{mu-N}
\ee 
The total force is written as a sum of
the external force including the driving  homogeneous electric force $e\m E_0$ and the Lorenz force in the external homogeneous magnetic field $\m B$,
$$\m F_0 = e \left( \m E_0 + \left[ \frac{\m v}{c} \times \m B\right]\right),$$ 
and the inhomogeneity-induced correction 
$\delta \m F $ that depends on the electrostatics of the problem. For the gated case, when interaction is screened beyond the gate-to-channel distance $d$, this correction reads  
\be 
\delta \m F = - \frac{e^2 \nabla N}{\cal C},
\label{dF-gated}
\ee  
where    ${\cal C}=\varepsilon/4\pi d$   is the gate-to-channel capacitance per unit area, and  $\varepsilon$ is the dielectric constant. For comparison, in the case of unscreened long-range Coulomb interaction, the correction reads:
$$ \delta \m F = -e^2 \nabla \int d^2\m r^\prime \frac{N(\m r^\prime )  }{\varepsilon |\m r-\m r^\prime|} .$$

The term $N^{-1} \nabla W$ on the r.h.s. of Eq.~\eqref{dv-dt} represents the thermoelectric force, where   
$$W=W(N,T)= \int_0^\infty  d\epsilon \  \epsilon\  f_F (\epsilon)$$
is the density of energy in the frame moving with flow. Here, $f_F= 1/\exp[ (\epsilon - \mu)/T +1]$ is the  Fermi function and 
$\mu$ should be expressed via $N$ by using Eq.~\eqref{mu-N}. In the limiting cases, this function is given by  
\be
W \approx \left\{\begin{array}{c} \displaystyle
     \frac{N^2}{2\nu}+ \frac{\pi^2 \nu T^2}{6},\quad{\rm for}\quad  \nu T \ll N,~ \\[0.4cm]
\displaystyle
    T N, \quad{\rm for}\quad \nu T \gg N.
 \end{array} 
 \right.
\label{W-mu}
\ee
The temperature balance  equation \eqref{dC-dt}  contains the heat capacity defined as
\BEA
C&=&C(N,T)= (\p W /\p T)_{N=\text{const}}
\nonumber
\\
&\approx& \left\{\begin{array}{c} \displaystyle
      \frac{\pi^2 \nu T}{3},\quad{\rm for}\quad \nu T \ll N,~  \\[0.4cm]
\displaystyle
    N, \quad{\rm for}\quad \nu T \gg N.
 \end{array} \right.
\label{C-N}
\EEA
The term $-\gamma (T-T_0)$ on the r.h.s. of Eq.~\eqref{dC-dt}  governs the heat transfer from the electron system with temperature $T$ to the phonon bath with temperature $T_0$.  
Strictly speaking, the phonon  temperature $T_0$ also depends on $\m r$.  Throughout  most of the paper we neglect this dependence assuming that $T_0$ is fixed by fast heat exchange with the substrate.  A brief discussion of the dependence $T_0(\m r)$ is presented in Sec.~\ref{sec:phonon_temp}. 

\subsection{ Model of inhomogeneity} 
\label{sec:constriction}

In this paper, we consider a simple model of inhomogeneity, which allows for an analytical solution. First of all, we assume that the sample inhomogeneity is completely governed by the spatial dependence of the momentum relaxation time, $\tau(\m r)$,  within a certain constricted area, while other coupling parameters and the external fields are position-independent. We limit ourselves to the weak-inhomogeneity limit, for which the dimensionless parameter  
\begin{equation}
\xi (\m r) = \frac{\tau_{\infty}}{\tau (\m r )} - 1 
\label{eq:xi_define}
\end{equation}
is  small, $\xi \ll 1$, and search for solutions of Eqs.~\eqref{dN-dt}, \eqref{dv-dt}, and  \eqref{dC-dt} perturbatively, to the leading order in $\xi$. 
In Eq.~\eqref{eq:xi_define}, $\tau_{\infty}$ is the  momentum relaxation time 
far away from the constricted area.

Recently, the quasi-1D case was analyzed in Ref.~\cite{Tikhonov2019}, where the temperature distribution of an infinitely long stripe-shape sample was found. In the quasi-1D geometry, the parameter $\xi$ only depends on the coordinate $x$ along the strip and is nonzero only in a limited area  (see Fig.~\ref{fig:model}a).
The asymmetry of temperature distribution in a variety of different overheating regimes, both in the hydrodynamic  and in the impurity-dominated (the so-called drift-diffusion) cases, was predicted
[see Fig.~\ref{fig:model}(b)]. 
This asymmetry is captured neither by the conventional local Joule heating approximation, nor a more advanced approach of Ref.~\cite{RokniLevinson1995}), and reveals itself in the regime of sufficiently strong non-equilibrium.  
             
In this paper, we consider the hydrodynamic heat transport in higher dimensionalities,  mostly focusing on 2D cases. We study the heat flow through a spherical-symmetric constriction  
characterized by the transport relaxation time $\tau_0 $ at the center of the constriction  and the radius $L$. For definiteness, when illustrating our results we use the Gaussian shape of the inhomogeneity:
\begin{equation}
\xi(\m r) = \left( \frac{\tau_{\infty}}{\tau_0} - 1 \right) \exp\left( - \frac{r^2}{L^2} \right).
\label{eq:circular_constriction}
\end{equation}

\section{Current-induced heating  }
\label{sec:current}

\subsection{Homogeneous heating}
\label{sec:homo}

We start our analysis with the homogeneous case, $\tau(\m r)=\tau_\infty$, $\xi \equiv 0$.  
We fix the electric field at $|\m r| \to \infty$ and assume that the magnetic field is parallel to the $z$ axis: $\m B = - B \m e_z.$ 
From Eqs.~\eqref{dN-dt}, \eqref{dv-dt}, and \eqref{dC-dt} we find       
the homogeneous velocity $\m v_\infty $ 
and the homogeneous temperature $T_{\infty}$
\begin{equation}
\m{v}_\infty = \chi \frac{\m E_0 -  \beta~ \m E_0 \times \m e_z  }{1+ \beta^2},
\qquad T_\infty  = T_0 + 
\frac{m v_\infty^2}{\tau_\infty \gamma}. 
\label{eq:zero_b_homogeneous}
\end{equation}
Here, $\chi = e \tau_\infty/m$,  
$\beta= \omega_c \tau_\infty= \tan \theta$,  $\omega_c=eB/mc$,  $\theta$ is the Hall angle, 
\be
v_\infty=\frac{\chi E_0}{\sqrt{1+\beta^2}}= \chi E_\parallel 
\label{v-inf}
\ee
is the absolute value of velocity, and $E_\parallel=E_0 \cos \theta $   is the projection of the electric field on the direction of electric current.
Following  Ref.~\cite{Tikhonov2019} we define the dimensionless overheating parameter  
\begin{equation}
\alpha = 1 - \frac{T_0}{T_\infty} = \left( 1 + \frac{m\gamma T_0}{e^2 E_\parallel^2 \tau_{\infty}} \right)^{-1},
\label{eq:over_heating}
\end{equation}
which scales quadratically with a weak electric field and  saturates  at $\alpha = 1$
in strong fields.

From Eqs.~\eqref{eq:zero_b_homogeneous} and \eqref{v-inf}, we realize that the application of the magnetic field for a fixed direction of external electric field simply introduces a rotation of all the profiles by the Hall angle $\theta$. In what follows, we thus focus only on the case of zero magnetic field.

\subsection{Inhomogeneous  heating}

In this section, we study the spatial profiles
of the hydrodynamic variables (density, hydrodynamic velocity, and temperature) after the introduction of a weak inhomogeneity.
For $B=0$, $\theta=\beta=0,$
$\m v_{\infty}=\mu \m E_0,$ and $ E_\parallel =E_0.$
Below, we consider the two distinct interacting models.

\subsubsection{Gated  2D liquid}
\label{sec:circular_constriction}

Let us assume that the electron-electron interaction is screened by a gate, so that $\delta \m F$ is given by Eq.~\eqref{dF-gated} . We introduce small  inhomogeneity-induced   corrections $\delta n$, $ \delta  \m v$, and  $\delta T$, which are proportional to $\xi$:   
\begin{align}
N& = N_\infty (1+\delta n),
\\
\m v&=\m v_\infty + \delta \m v,
\\
T&=T_\infty + \delta T .
\end{align}
Linearizing Eqs.~\eqref{dN-dt}, \eqref{dv-dt}, and  \eqref{dC-dt} with respect to $\xi$  and taking Fourier transform we get (for the Fourier transformed quantities):
\begin{align}
&\ \m q  \m v_{\infty} \delta n_{\m q}+ \m q \delta \m v_{\m q}  =0,
\label{dn}
\\
&\left(\! \frac{1}{\tau_{\infty}} +i \m q \m v_{\infty}\!\right)\delta \m v_\m q\! +\! i \m q \left (\!  s^2 \delta n_\m q\!  +\! \frac{\delta T_\m q}{M} \!\right) =-\frac{\m v_\infty}{\tau_\infty} \xi_\m q,
\label{dv}
\\
&\left (\! \frac{\gamma}{c_\infty} \!  +\! i \m q   \m v_{ \infty}\! \right )\! \delta T_\m q \!  +\! 
\left(\!i \m q T_\infty\! -\! \frac{2M \m v_\infty }{\tau_{\infty}}\!\right)\! \delta \m v_\m q \!     =\! \frac{M v_\infty^2}{\tau_\infty} \xi_\m q.
\label{dT}
\end{align}
Here
\be
s^2=\frac{e^2 N_\infty}{m \cal C} + \frac{1}{m}\left( \frac{\p W}{\p N} \right)_{T=\text{const}}
\ee
is the plasma wave velocity, with the second term representing contribution from the Fermi-liquid velocity. We also define
\be
M= \frac{m}{c_\infty}
\ee
as the ``thermal'' mass, and
\be
c_\infty=\frac{C_\infty}{N_{\infty}}= \frac{1}{N_{\infty}} \left(\frac{\p W}{\p T }\right)_{N=\text{const}}
\ee
is the specific heat capacity. 

\subsubsection{Weakly compressible liquid}
\label{weakly comppressible}
General solutions of Eqs. \eqref{dn}, \eqref{dv},  and \eqref{dT}  are  rather cumbersome. Below we present solutions for the most interesting case, where the liquid is almost incompressible. This case is realized for strong electron-electron interaction, when the
plasma wave velocity is large:
\be
s \gg \sqrt{\frac{T_\infty}{ M} }, \quad s \gg v_\infty.
\label{s-gg-T}
\ee
We introduce two characteristic lengths:
the elastic drift length determined by the scattering off disorder, 
\be
 l=v_\infty \tau_\infty,  
\ee
and the inelastic drift length~\cite{Tikhonov2019} characterizing the electron-phonon scattering,
  \be
l_{\text{in}} = \frac{v_\infty  c_\infty} {\gamma}. 
\ee
We further assume that the elastic drift length is the shortest lengthscale,
\be
L \gg l, \qquad l_{\rm in} \gg l,
\label{ineq}
\ee
while the relation between the constriction size, $L$, and  $l_{\rm in}$ can be arbitrary.
 
From  Eq. ~\eqref{dv}  one can conclude that, in the limit $s \to \infty$, the  correction to the electron density is small, $\delta n \propto 1/s^2$.  Then, from  Eq.~\eqref{dn}, 
we find that $ \m q \delta \m v_\m q \propto 1/s^2$,   so that the electron liquid is almost incompressible: $\text{div} \delta \m v \to 0.$ As a consequence, to the zeroth order with respect to $1/s^2$,  one can replace Eq.~\eqref{dn} with 
\be
\m q \delta \m v_{\m q}=0.
\label{dv-incomp}
\ee
Hence, the velocity correction in the momentum space is perpendicular  to $\m q$:   
$$\delta \m v_{\m q} \propto \m t_\m q ,$$ 
where $ \m t_\m q= [\m e_z \times \m q/q] $, and $\m e_z$ is the unit vector in the $z$ direction. This correction can be found by  taking the scalar product of 
$\m t_ \m q $ and Eq.~\eqref{dv}. 
Substituting $\delta \m v_\m q$ into  Eq.~\eqref{dT} and neglecting the small term  $\m q\delta \m v_\m q $, one can find $\delta T_\m q$. 
Having in mind Eq.~\eqref{ineq}, we finally arrive at the following expressions for the velocity and temperature corrections:
\begin{align} 
&\delta \m v_\m q= - \xi_\m q  \m t_\m q (\m t_\m q \m v_\infty ),
\label{dv-final}
\\
&\delta T_\m q= \frac{M \xi_\m q v_\infty }{\tau_\infty}
\frac{q_{\parallel}^2-q_\perp^2}{q^2 (i q_{\parallel} + 1/l_{\rm in})}.
\label{dT-final}
\end{align}
Here, $q_{\parallel}$ and $q_{\perp}$ are, respectively,  the   parallel  and perpendicular components  of the vector $\m q$ with 
respect to the drift velocity $\m v_\infty$.

A correction to the concentration arises only in the first order with respect to $1/s^2$. 
It can be obtained by taking the scalar product of Eq.~\eqref{dv} and $\m q$: 
\be
\delta n_{\m q}= \frac{\xi_{\m q} v_\infty}{s^2 \tau_{\infty}   q^2}
\left( i  q_\parallel - 
\frac{ q_\parallel^2 - q_\perp^2}
{ i  q_\parallel+ 1/l_{\rm in} }
\right).
\ee
Importantly, although $\delta n_\m q $ approaches zero for large $s$, one cannot neglect $\delta n_\m q$ from the very beginning, because $ \delta F_\m q \propto s^2 \delta n_\m q$ remains finite for $s \to \infty$.

We see that the temperature distribution in the momentum space is  described (up to a constant coefficient) by the product of $\xi_\m q$ and the heating kernel 
\be
K(\m q)=\frac{q_\parallel^2-q_\perp^2}{q_\parallel^2+q_\perp^2} ~ \frac{1}{ i q_\parallel  + 1/l_{\rm in}   }.
\label{Kq}
\ee
It is worth stressing that this equation is derived in the incompressible limit corresponding to very strong interactions and, therefore, does not contain
any specific feature of the 2D system.  
One can easily show that this form of the heating kernel universally appears in other dimensions. In particular,  for quasi-1D strips parallel to the $x$-axis, where 
$ \xi=\xi (x),$  the transverse wave vector equals zero, $q_\parallel=0$, and we arrive at equations derived previously in Ref.~\cite{Tikhonov2019}:   
$$K(q_x)= 1/(i q_x+ 1/l_{\rm in}),$$ 
see Eq.~(27) in that work.  
By using the same calculations as presented above, one finds that for a 3D case, the heating kernel is also given by Eq.~\eqref{Kq}, with
$q_\perp^2=q_y^2+q_z^2$ (for $\m v_\infty $ parallel to $x$ axis).

\subsubsection{Landauer-dipole structure of the heating kernel}
\label{Landauer-dip}

Direct calculation of the Fourier transform of \eqref{Kq} yields
\be
K(\m r)= \int \limits _0^\infty ds e^{-s/l_{\rm in}} Q_D(x_\parallel- s, \m r_{\perp}),
\label{K}
\ee
where $Q_D(\m r)$ is equivalent to the  ``electric field'' of a Landauer dipole \cite{Landauer}, which can be written in the universal form for all dimensions:
\be
Q_D(\m r)= C_{D} \frac{\p}{\p x_\parallel} \left( \frac{x_\parallel}{r^D}\right),\qquad D=1,2,3,
\ee
where $C_1=1/2,~C_2=1/\pi,~C_3=1/2\pi$ for 1D, 2D and 3D cases.   
Choosing $x$ axis along the  $\m v_ \infty     $ direction, we get the explicit expressions for Landauer dipoles in the heating kernel: 
\be
Q_D(\m r) =\left \{ \begin{array}{cc} \displaystyle
      \delta(x),~ \quad &\text{for}\quad D=1,  
     \\[0.1cm]
      \displaystyle \frac{y^2-x^2}{\pi (x^2+y^2)^2},~\quad &\text{for}\quad D=2,
     \\[0.2cm]
     \displaystyle  \frac{y^2+z^2-2 x^2}{2\pi (x^2+y^2 +z^2)^{5/2}},~\quad & \text{for}\quad D=3.
\end{array}   \right. 
\label{QD}
\ee
These functions obey the following property: $$Q_1(x)=\int dy\, Q_2(x,y)=\int dy\,dz\, Q_3(x,y,z).$$ 
In order to illustrate the physics behind the Landauer-dipole-like temperature distributions, we present in Appendix~\ref{app:multi_conduct} the discussion of the Joule heat distribution in a simple two-component model within the conventional theory of local Joule dissipation.

As seen from Eq.~\eqref{K}, the heating kernel is given by  a  Landauer dipole that is spatially-shifted  at the distance
$\sim l_{\rm in}$.   
This shift induces the asymmetry of the Landauer dipole along the direction of the current. Physically, the asymmetry stems from ``convection'' described by the term 
$\text{div}( \m v T)$ in the heat balance equation which is absent in the theory of local dissipation and in the weak-drive theory of non-local heat transport of Ref.~\cite{RokniLevinson1995}.  Interestingly, the kernel Eq.~\eqref{K} remains finite in the limit $l_{\rm in} \to \infty$, which means that the temperature distribution in this case is fully determined by convection.

The correction to the electron temperature is given in the coordinate space by
\be
\delta T (\m r)= \frac{T_0}{l_{\rm in}} \frac{\alpha}{1-\alpha} \int d^D \m r^\prime K(\m r- \m r^\prime) \xi(\m r^\prime). 
\ee
The asymmetry of temperature distribution manifests itself at distances of the order of (or smaller than) the inelastic lengths. 
At larger distances, we get
\be
K(\m r) \approx l_{\rm in} Q_D(\m r ),\quad \text{for} \quad r\gg l_{\rm in}. 
\ee
Hence, away from the inhomgeneity center, for $r \gg \text {max} (L,l_{\rm in})$, the temperature distribution can be considered as symmetric and  given by the Landauer-dipole profile:
\be
\delta T (\m r) \approx \frac{\alpha T_0  }{ (1-\alpha)}  ~   Q_D(\m r)  \int \xi (\m r^\prime) d^D\m r^\prime,\quad \text{for} 
\quad r\to \infty.
\ee

For the case of the Gaussian constriction \eqref{eq:circular_constriction}, equations for temperature distribution valid for arbitrary relation between $r,~L$ and
$l_{\rm in}$ are derived in Appendix \ref{Gauss}. They can be written as follows.
\be
\delta T (\m r)=   A_D \int \limits _0^\infty \frac{ds}{l_{\rm in}} e^{-s/l_{\rm in}} \tilde Q_D(x- s, \m r_{\perp}),
\label{eq:temp_gaussian}
\ee
where
\be
A_D=\pi^{D/2} L^D \frac{\alpha }{1-\alpha}   \left(\frac{\tau_\infty}{\tau_0} -1\right) T_0  ,
\ee
and 
\be \label{tildeQ}
\begin{aligned} 
  \tilde Q_1  &= \frac{1}{\sqrt{\pi} L} e^{-x^2/L^2} ,
\\
 \tilde Q_2 &= \frac{y^2  - x^2 }{ \pi r^4}  \left[ 1-  e^{-r^2/L^2} (1+ r^2/L^2) \right], 
\\
  \tilde Q_3 &=  \frac{(y^2 +z^2 -2 x^2)~ \text{Erf} ( r/L) }{2 \pi r^5}
\\ 
&- \frac{e^{-r^2/L^2} [  (y^2+z^2)^2 -x^4 + L^2 (z^2+y^2-2 x^2) ] }{\pi^{3/2} L^3 r^4}.
\end{aligned}
\ee
Sending $L \to 0$ for fixed  $r$, we reproduce Eqs.~\eqref{QD}:  $\tilde Q_D \to Q_D$ for $L \to 0$.

Above, we have assumed that the elastic drift length $l$ is much smaller in comparison to the constriction size: $l \ll L$.  The obtained results can be straightforwardly generalized to the case of  arbitrary relation between $l$ and $L$. As follows from Eqs.~\eqref{dn}, \eqref{dv}, and \eqref{dT}, the general expression for the heating kernel, Eq.~\eqref{Kq},  becomes then
\be
K(\m q)=\left(\frac{q_\parallel^2-q_\perp^2}{q_\parallel^2+q_\perp^2} +  i q_\parallel l\right) ~ \frac{1}{i q_\parallel  + 1/l_{\rm in}}
\frac{1}{1+ i q_\parallel l }.
\label{Kq1}
\ee
In real space, the heating kernel is again expressed in terms of the Landauer-dipole field:
  \be
   \begin{aligned}
K(\m r)&= \frac{l_{\rm in}}{l_{\rm in}-l}  \int \limits _0^\infty d\rho \left(e^{-\rho/l_{\rm in}} -e^{-\rho/l}\right) Q_D(x_\parallel- \rho, \m r_{\perp})
\\
&+ \delta(\m r_\perp ) \theta (x_\parallel) \left(  e^{-x_\parallel/l} - \frac{l}{l_{\rm in} }e^{-x_\parallel /l_{\rm in}}\right).
\label{K1}
  \end{aligned}
  \ee
Clearly, this reproduces Eq.~\eqref{K} in the limit $l \to 0$. As another feature, similar to Eq.~\eqref{K},  the kernel Eq.~\eqref{K1} remains finite in the limit of diverging $l_{\rm in}$:
  \be
   \begin{aligned}
K(\m r)& \approx \int \limits _0^\infty d\rho \left(1 -e^{-\rho/l}\right) Q_D(x_\parallel- \rho, \m r_{\perp})  
\\
&+ \delta(\m r_\perp ) \theta (x_\parallel)   e^{-x_\parallel /l}, \quad \text{for} \quad l_{\rm in} \to \infty .
\label{K11}
  \end{aligned}
  \ee

\subsection{Spatial profiles of the electron temperature, velocity, and concentration}

\begin{figure}
  \centering
      \includegraphics[width= 1 \columnwidth]{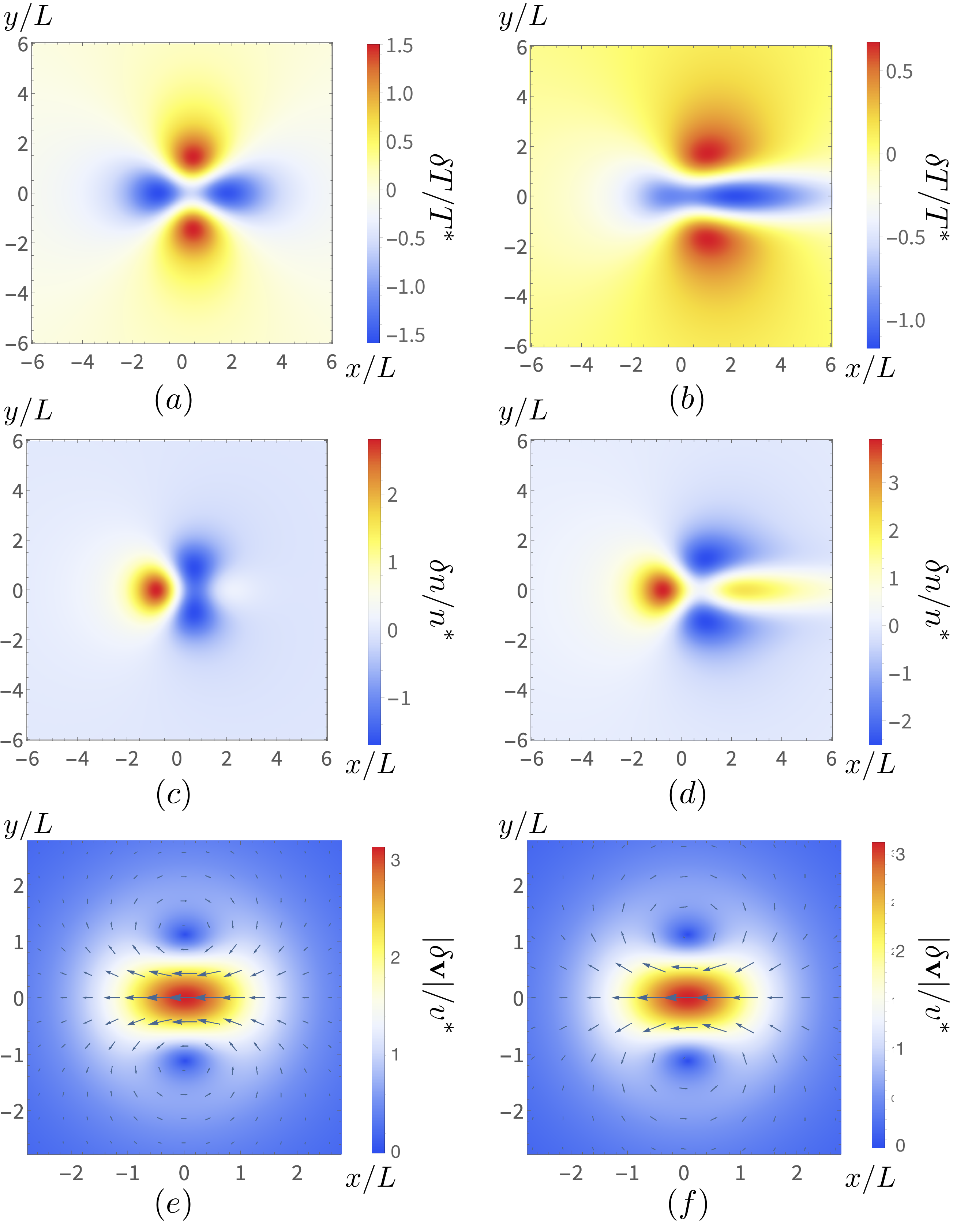}
  \caption{ Density plots of $\delta T/T_*$  in 2D systems with a Gaussian constriction,  Eq.~\eqref{eq:circular_constriction} in the incompressible limit  for
  $L = 2 l_{\text{in}}$ (a) and $L = 0.5 l_{\text{in}}$ (b), respectively; (c) and (d): corresponding density plots of $\delta n/n^*$; (e) and (f): vector 
  density plots of $\delta \m v/v^*$.  }
  \label{fig:large_ec}
\end{figure}

The spatial dependences of the temperature, as well as the velocity and concentration, are plotted  in Fig.~\ref{fig:large_ec} for a 2D system with a Gaussian  constriction [see  Eq.~\eqref{eq:circular_constriction}].  We assumed that the liquid is nearly incompressible ($s\to \infty$) and $l\ll l_{\rm in},~l\ll L. $   The temperature, velocity and concentration  are measured, respectively, in the  following units
\begin{align}
&\!\! T^{*}= \left(\frac{\tau_\infty}{\tau_0}-1 \right) \frac{\alpha}{1-\alpha}T_0 = \left(\frac{\tau_\infty}{\tau_0}-1 \right) (T_\infty-T_0),
\label{T*}
\\
&\!\!v^*=\left(\frac{\tau_\infty}{\tau_0}-1 \right) v_\infty,
\label{v*}
\\
&\!\!n^*= \left(\frac{\tau_\infty}{\tau_0}-1 \right)\frac{L}{l} \frac{v_\infty^2}{s^2}.
\label{n*}
\end{align}

The profiles of  $\delta T/T^*$,  $\delta v/v^*$, and $\delta n/n^*$ depend only on the relation between characteristic lengths of the problem, $L$, $l_{\rm in}$, and $l$. 
Figures~\ref{fig:large_ec}(a),(b) show the temperature plots $\delta T/T_*$
for $l \ll (L,l_{\rm in}) $ for the two different ratios between $L$ and $l_{\rm in}$: $L=2l_{\rm in}$ for Fig.~\ref{fig:large_ec}(a)  and   $L=0.5 l_{\rm in}$ for Fig.~\ref{fig:large_ec}(b). 

Panel (a) of Fig.~\ref{fig:large_ec} reproduces an almost symmetric Landauer-dipole structure, while panel (b) is much more asymmetric because of a larger value of $l_{\rm in}/L$. 
In both panels, the electron temperature is reduced along the current injecting direction (the $x$ direction), and enhanced in the direction perpendicular to the current (the $y$ direction). 
The Landauer-dipole structures are shifted by the inelastic drift length $l_{\text{in}}$, yielding the temperature asymmetry similar to the one predicted for the quasi-1D geometry~\cite{Tikhonov2019}.

The density plots of $\delta n/n^*$ for the same values of $L$ and $l_{\rm in}$ are shown in panels (c) and (d) of Fig.~\ref{fig:large_ec}. We observe that among these distributions the most asymmetric is the density profile, which emerges in the limit of nearly incompressible fluid only for finite values of $1/s^2$, in contrast to the temperature and hydrodynamic velocity. The asymmetry of the temperature profiles corresponds to a similar feature of the density distributions: particles tend to gather in front of the constricted area, where a larger impurity scattering rate is present.      

The appearance of the Landauer-dipole structure in the temperature distribution can be  understood based on the vector density plots for the velocity distribution $\delta \m v/v^*$ that are shown in panels (e) and (f) of Fig.~\ref{fig:large_ec}. One sees that particles  tend to detour around the constricted area. Heat current, carried by these detoured particles, thus leads to the hot spots on the $y$ axis, which are clearly seen in panels (a) and (b).

\section{The viscous case}
\label{viscosity}
So far, we have considered hydrodynamics of an ideal electron fluid characterized by zero viscosity
(taking the limit $\tau_\text{ee}\to 0$).
In the presence of a finite viscosity $\eta$, expressions for the velocity and temperature corrections are modified as follows:
\begin{align}
&\delta \m v_{\m q}= - \frac{\xi_{\m q} \m t_{\m q} (\m t_{\m q} \m v_{\infty})}{1 +i q_\parallel l + \kappa q^2 }
\label{eq:viscous_velocity}
\\
&   \delta T_\m q=   \frac{M \xi_\m q v_\infty }{\tau_\infty}
\frac{1}{ i q_{\parallel} + 1/l_{\rm in}} \left(1- \frac{2 q_\perp^2}{q^2} \frac{1}{1+ i q_\parallel l +\kappa q^2}\right).
\label{eq:viscous_temp}
\end{align}
Here, 
\be
\kappa= \eta \tau_{\infty} =l_\kappa^2,
\ee
where $l_\kappa=\sqrt{\eta \tau_\infty}$ is a  new length scale, arising in the presence of viscosity. It is worth noting that viscosity by itself gives a direct contribution to the dissipation \cite{dau6}, so that the term 
\begin{equation}
P_{\text{vis}} =  \eta  \iint dx dy  \left[ \left( \partial_x v_x - \partial_y v_y \right)^2 + \left( \partial_x  v_y + \partial_y v_x \right)^2  \right],
\label{eq:local_dissipation}
\end{equation}
appears on the r.h.s. of the heat balance equation, Eq.~\eqref{dC-dt}.
Here $v_x$ and $v_y$ are the spatially dependent velocities along the $x$ and $y$ directions, respectively. As seen from Eq.~\eqref{eq:local_dissipation}, the direct viscous dissipation is present only in the  second order in the weak inhomogeneity $\xi$. Hence, in the linear approximation with respect to $\xi$, the viscosity affects the temperature distribution indirectly by changing velocity distribution, thus modifying the impurity dissipation $mv^2/\tau$.  Depending on the relation between four characteristic lengths:
 $L$, $l$, $l_{\rm in}$, and $l_\kappa$, various regimes of dissipation are possible. However, the main effect of viscosity is quite simple: with increasing $l_\kappa$, the Landauer-dipole structure is suppressed. It makes sense, therefore, to start with the discussion of the high-viscosity limit.

\subsection{Strong viscosity, $l_\kappa \to \infty$}
Let us assume that $l_\kappa $ is larger than all other length scales in the problem. 
Interestingly, even in the limit $l_\kappa \to \infty$, there is a finite temperature correction, which can be found by neglecting the last term of Eq.~\eqref{eq:viscous_temp}. Performing the Fourier transformation, we get 
\be
\delta T = \frac{M v_{\infty}^2 }{l}\int \limits _{-\infty } ^{x_{\parallel}} dx_{\parallel}^\prime ~\xi(\m r^\prime) e^{-(x_{\parallel}-x_{\parallel}^\prime)/l_{\rm in}}.
\label{eq:strong_vis_temp}
\ee
Let us comment on the physics behind this expression. For infinitely large viscosity, all velocity gradients are suppressed, so that $\m v \equiv \m v_{\infty}$. In this limit, the heat balance equation is dramatically simplified  
\be
 \frac{d \delta T}{dx_\parallel}= \frac{M v_{\infty}^2 ~\xi(\m r)}{l}  -  \frac{\delta T}{ l_{\rm in}}.
\label{eq:strong_vis_kinetic_eq}
\ee
Physically, this equation means that the heating is effectively one dimensional, regardless of the system dimension. Integrating Eq.~\eqref{eq:strong_vis_kinetic_eq}, we reproduce Eq.~\eqref{eq:strong_vis_temp}.  
We also notice that Eqs.~\eqref{eq:strong_vis_temp} and \eqref{eq:strong_vis_kinetic_eq} do not require linearization and are valid for an arbitrary relation between $\delta T$ and $T_\infty$.

 \begin{figure}
  \centering
      \includegraphics[width=1 \columnwidth]{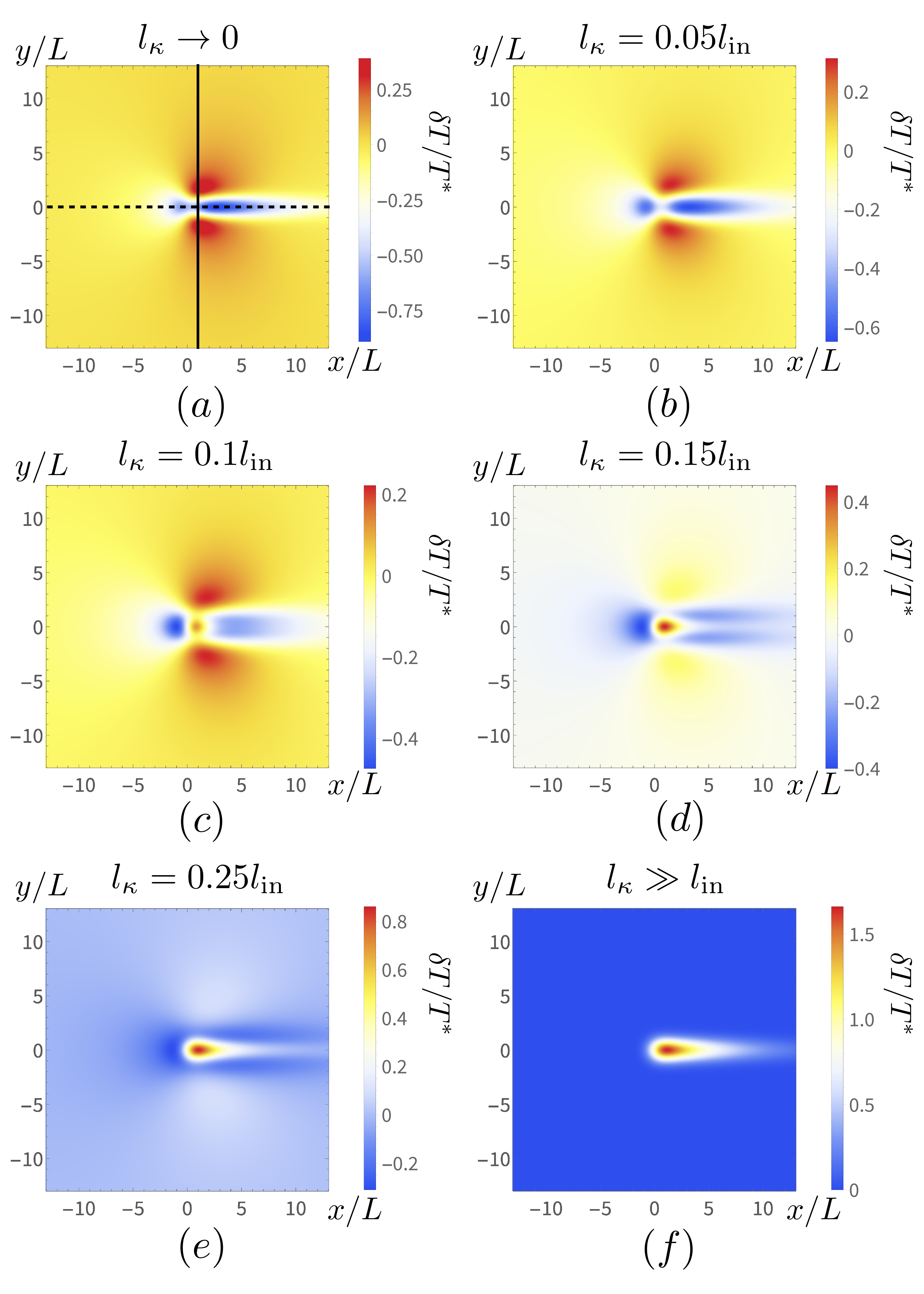}
  \caption{Evolution of the 2D temperature  profiles for fixed $L$ and $l_{\rm in}$, $L = 0.2 l_{\text{in}}$, with increasing $l_\kappa$:  (a) the limit of negligible viscosity,  $l_{\kappa} = 0.001 l_{\text{in}}$; (b) small viscosity,   $l_{\kappa} = 0.05 l_{\text{in}}$, which is insufficient to suppress the Landauer-dipole feature;  (c) ``critical'' viscosity, $l_{\kappa}  = 0.1 l_{\text{in}}$,  at which a hot spot  appears  in the center of the constriction, which increases  with further increasing viscosity, as shown in 
  (d) for $l_{\kappa} = 0.15 l_{\text{in}}$   and (e) for  $l_\kappa = 0.25 l_{\text{in}}$, gradually suppressing the Landauer-dipole structure; (f) strong viscosity, $l_{\kappa} = 10 l_{\text{in}}$: the Landauer-dipole feature is suppressed.}
  \label{fig:vis_temp_2d}
\end{figure}

\subsection{Evolution of the temperature profile with viscosity}

\begin{figure}
  \centering
      \includegraphics[width=1 \columnwidth]
      {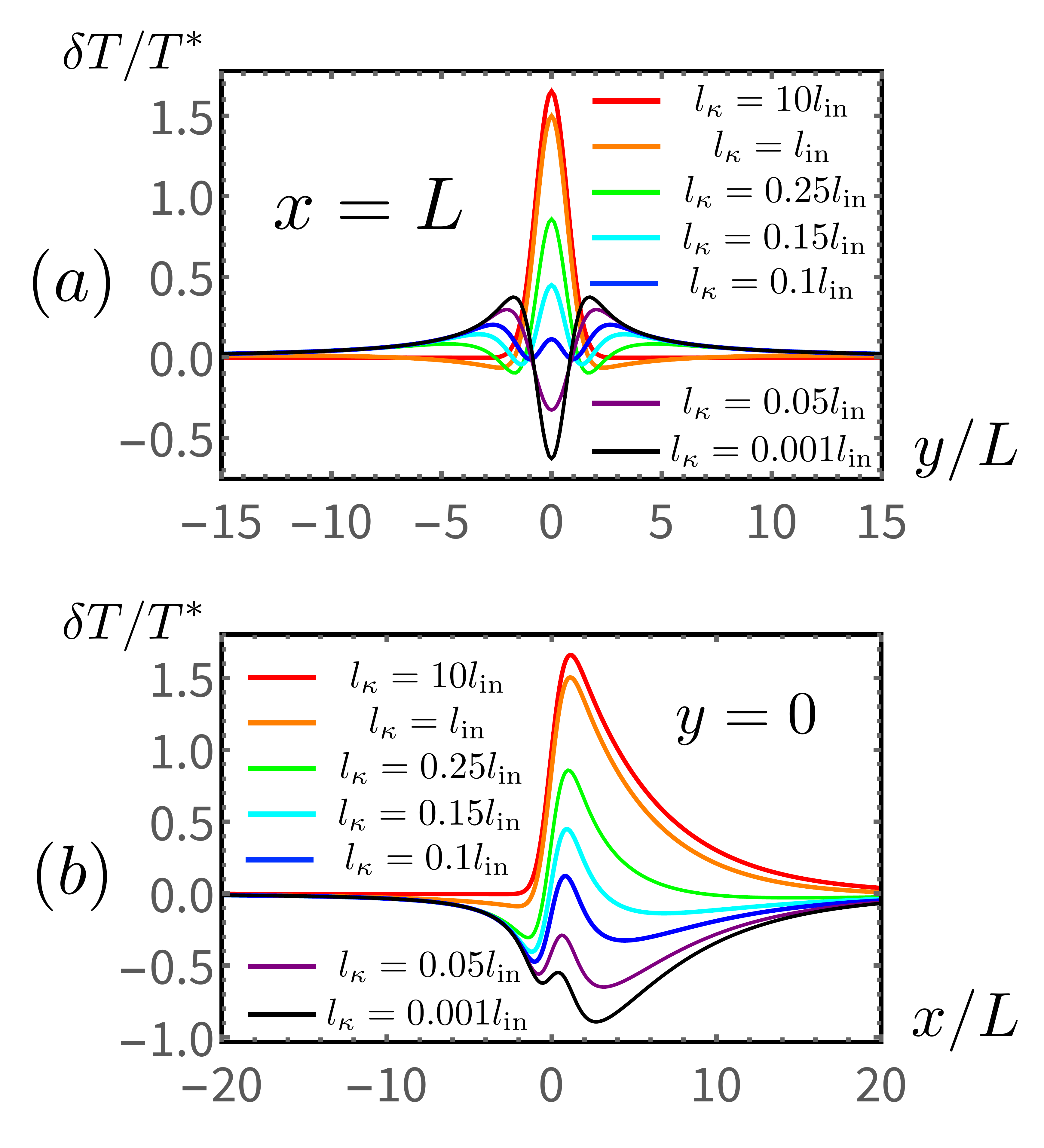}
  \caption{Evolution of the temperature profiles of the  cross-sections at (a) $x=L$ and (b) $y = 0$, corresponding to the solid and dashed lines in Fig.\,\ref{fig:vis_temp_2d}(a). Parameters are the same as in Fig.\,\ref{fig:vis_temp_2d}.}
  \label{fig:vis_temp_1d}
\end{figure}

In Figs.~\ref{fig:vis_temp_2d} and \ref{fig:vis_temp_1d}, we illustrate the evolution of the temperature distribution with increasing viscosity. We fix $L$ and 
$l_{\rm in}$ such that the inelastic drift length is larger than the size of the constriction, $L = 0.2 l_{\text{in}}$, and, thus, asymmetry in the temperature distribution  is sufficiently pronounced. At the same time, the Landauer-dipole structure remains apparent for zero and very small viscosity,  as shown in Figs.~\ref{fig:vis_temp_2d}(a) and (b). 
When $l_\kappa$ becomes the order of $L$, a hot spot appears  in the center of the constriction, Figs.~\ref{fig:vis_temp_2d}(c), whose  amplitude increases with increasing viscosity, see Figs.~\ref{fig:vis_temp_2d}(d), (e), and (f),  while,  simultaneously, the Landauer-dipole structure becomes suppressed. At very large values of $l_\kappa$,   Figs.~\ref{fig:vis_temp_2d}(f), the  Landauer-dipole feature is fully suppressed, and the temperature distribution is very well described by Eq.~\eqref{eq:strong_vis_temp}. 

\begin{figure}
  \centering
      \includegraphics[width=1 \columnwidth]{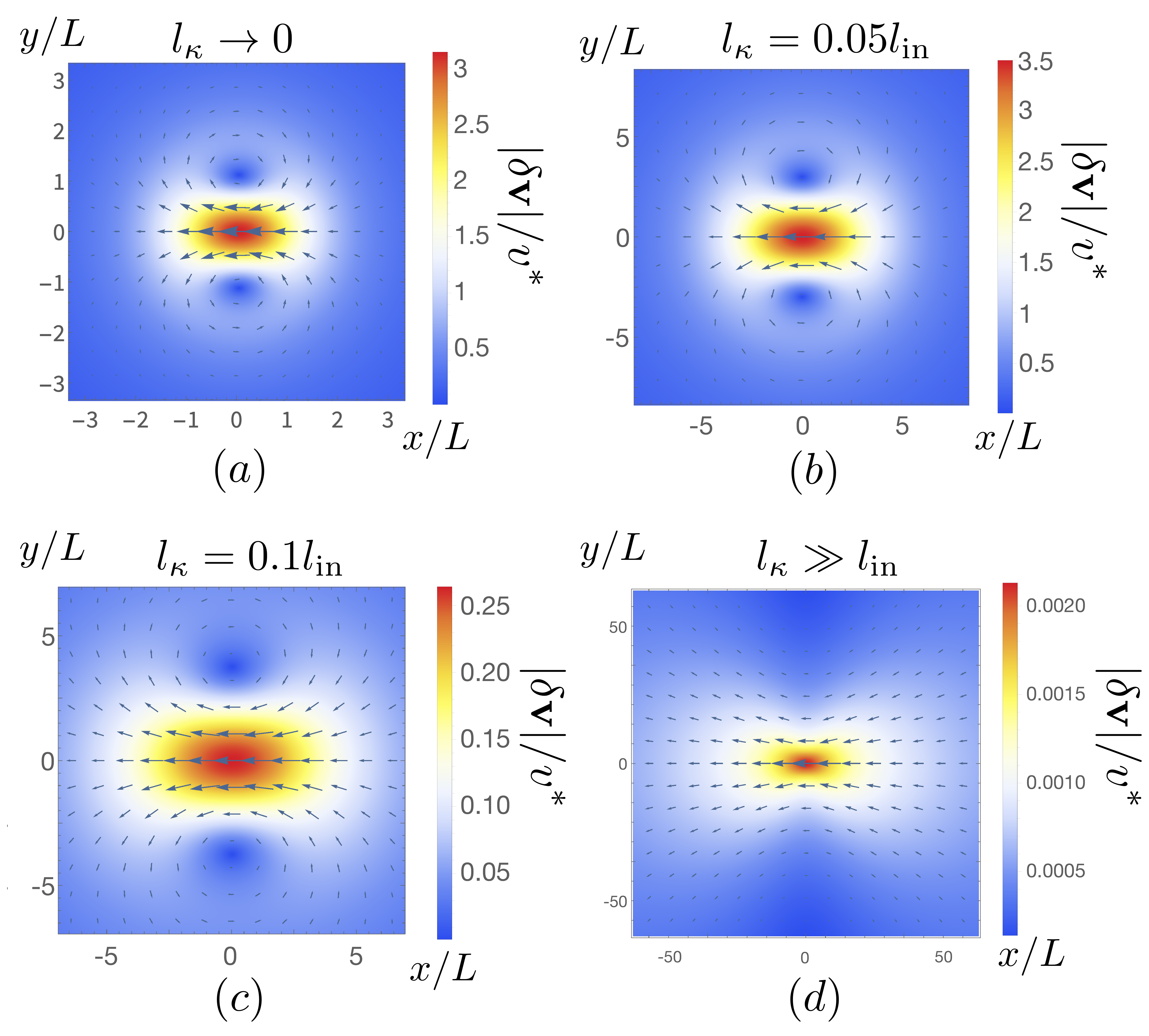}
  \caption{Evolution of the  velocity profiles with increasing viscosity.  Panels (a), (b), (c) and (d) correspond to panels  (a), (b), (c) and (d) in Fig.\,\ref{fig:vis_temp_2d}, respectively. 
  }
  \label{fig:vis_velocity}
\end{figure}

We also plot in Figs.~\ref{fig:vis_temp_1d}(a), (b) the cross-sections of the temperature  distribution  along the lines $x = L$ and $y=0$, respectively. These plots clearly demonstrate the suppression of the Landauer dipoles and the formation of the hot spot, which is symmetric in the $y$ direction but asymmetric in the $x$ direction.  Evolution of the velocity profiles with viscosity is shown in Fig.~\ref{fig:vis_velocity} for the same values of parameters as in Fig.~\ref{fig:vis_temp_2d}   

\begin{figure}
  \centering
      \includegraphics[width=1 \columnwidth]{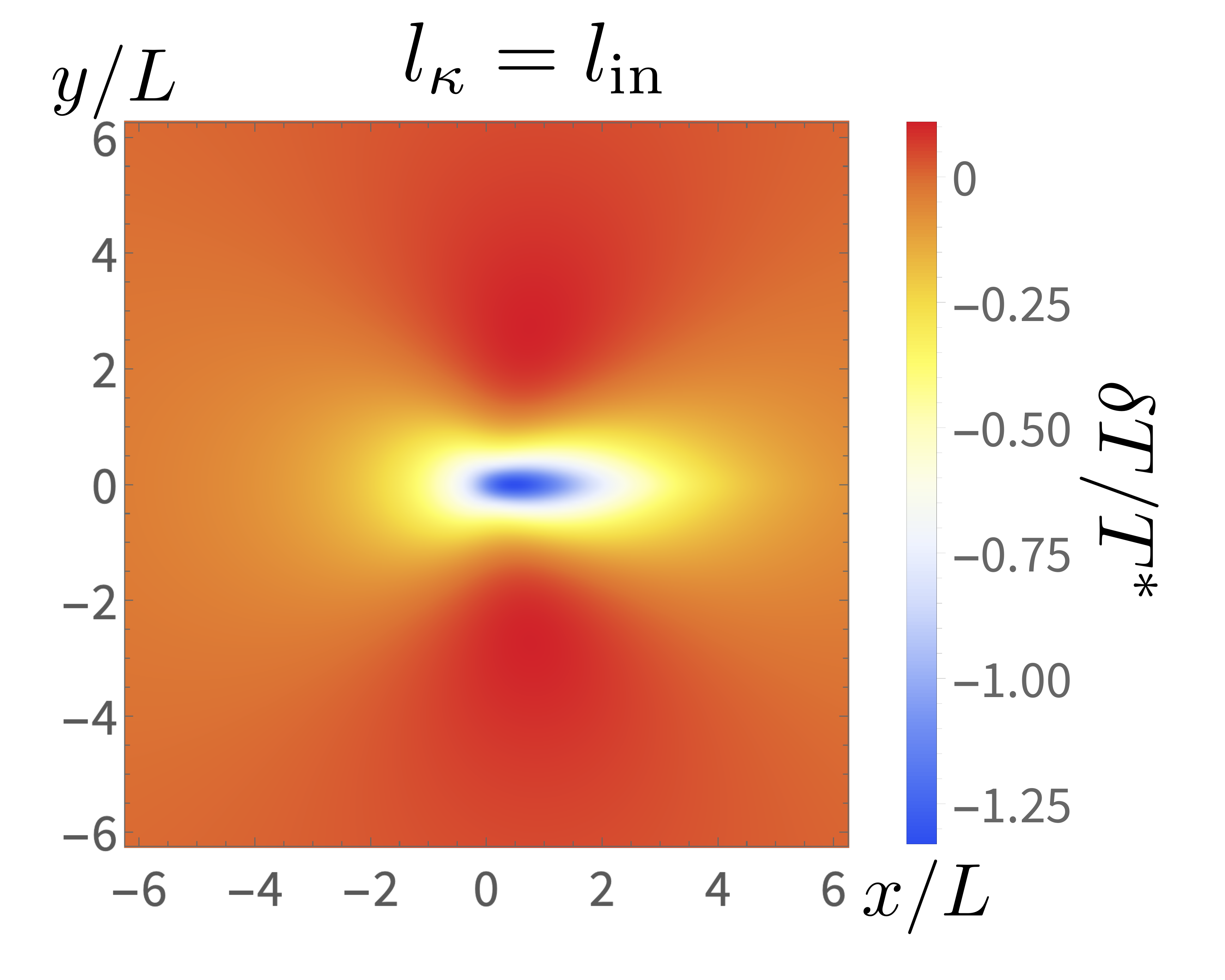}
  \caption{The temperature profile after the removal of the ``hot spot'' at the constriction position, see  Eq.~\eqref{eq:strong_vis_temp}.}
  \label{fig:ld_large_vis}
\end{figure}

We reiterate that, to the leading order in $\xi$, viscosity does not introduce extra dissipation, but instead redistributes it.
To see this point, in Fig.~\ref{fig:ld_large_vis} we present the temperature profile ($l_\kappa = l_{\text{in}} \gg L$) obtained when we keep only the second term of Eq.~\eqref{eq:viscous_temp} in the parentheses [i.e, after removing the contribution of Eq.~\eqref{eq:strong_vis_temp}].
Apparently, the Landauer-dipole structure survives in the strong-viscosity limit, although it is strongly reduced and obscured by the ``hot spot'' at the constriction.

  \section{Phonon temperature distribution}
\label{sec:phonon_temp}

In previous sections, we have discussed the
distribution of the electron temperature. 
However, experimentally, the electron temperature profiles are hard to measure directly. On the other hand,
the phonon temperature can be measured at a very high resolution by means of the tSOT technique~ \cite{FinklerNanoLett10,Vasyukov,HalbertalNat16,HalbertalScience17,Zeldov2019}.
Using the results obtained above, we obtain the phonon temperature, restricting ourselves to the discussion of the 2D case.

\begin{figure}
  \centering
      \includegraphics[width=1 \columnwidth]{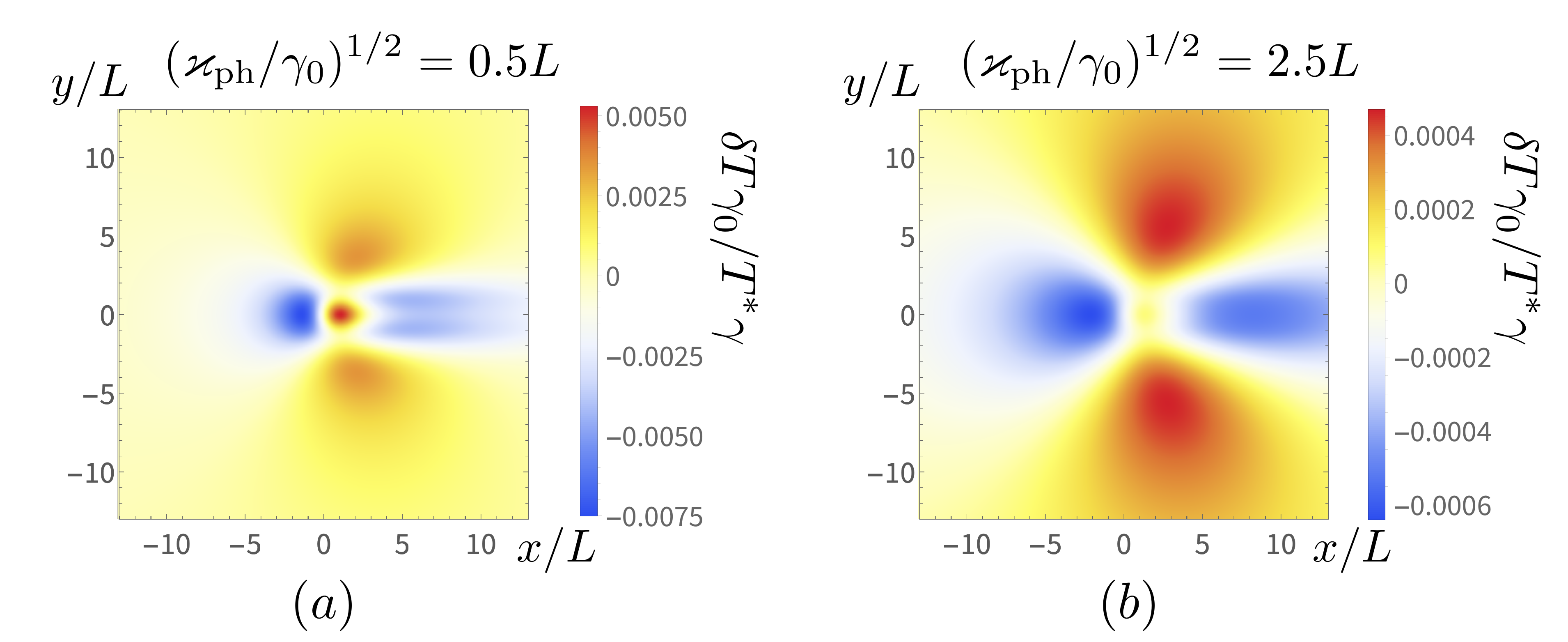}
  \caption{The phonon temperature near the constriction that corresponds to the electron temperature of Fig.~\ref{fig:vis_temp_2d}(d). The phonon temperature profile reproduces the major features of the electron one when $\varkappa_{\text{ph}}$ is small (a). On the contrary, the ``hot spot'' disappears when $\varkappa_{\text{ph}}$ becomes large enough (b).}
  \label{fig:phonon_temperature}
\end{figure}
 
First, we note that the heat exchange rate between phonons and the substrate, $\gamma_0$,  is typically several orders of magnitude larger than the electron-phonon heat transfer rate $\gamma$ (see experimental measurements  \cite{ChenAPL09,PerssonEPL2010,LiJoPD17} and estimates in Ref.~\cite{TikhononvPRB18}).
In the homogeneous case we have
\begin{equation}
\gamma (T_{\rm el} - T_{\rm ph} )= \gamma_0(T_{\rm ph} -T_0),
\end{equation}
where  $T_0$ is the substrate  temperature. For $\gamma_0 \gg \gamma$, we have $T_{\rm ph} \approx T_0$.
It is worth noting that this estimate justifies the approach used in the previous sections, where, for determining the electron temperature, we assumed that the lattice temperature is a constant.

Small deviations of the lattice temperature from $T_0$  can be found from the phonon heat balance equation
\begin{equation}
 -\varkappa_{\text{ph}} \Delta T_{\text{ph}}= \gamma ( T_{\text{e}} - T_{\text{ph}} ) - \gamma_0 (T_{\text{ph}} - T_0 ),
\label{eq:differential_set}
\end{equation}
where $ \varkappa_{\text{ph}}$ is the phonon thermal conductivity.  Introducing 
$$\delta T_{\rm ph} =T_{\rm ph}-T_0,$$ 
and making use of the inequality $\gamma\ll \gamma_0$, we find 
\be
\delta T_{\rm ph} ^{\m q}=
\frac{\gamma}{\gamma_0 + \varkappa_{\text{ph}}q^2} \delta T_{\m q},
\label{dTph}
\ee
where $\delta T_{\m q}$ is the inhomogeneity-induced correction to the electronic  temperature calculated above. 
We see that for  
$$q < q_0= \sqrt{\gamma_0/\varkappa_{\rm ph}},$$ the phonon temperature profile coincides with the electronic one up to a small factor $\gamma/\gamma_0$.   
Very sharp gradiends of the electronic temperature with $q > q_0$ are suppressed by the phonon heat conductivity.  
In the coordinate representation,  Eq.~\eqref{dTph} becomes
\be
\delta T_{\rm ph} (\m r)
=\int d\m r^ \prime K(\m r -\m r^\prime) \delta T(\m r ^\prime ),
\ee
where 
\be
K(\m r )=
\int \frac{d^2\m q ~e^{i \m q \m r}}{(2\pi)^2} \frac{ \gamma}{\gamma_0 +\varkappa_{\rm ph} q^2} = \frac{\gamma}{2\pi \varkappa_{\rm ph}} 
K_0(q_0 r),
\ee
and $K_0$ is the MacDonald function. 
Evolution of the phonon temperature distribution with increasing phonon heat conductivity is illustrated  in Fig.~\ref{fig:phonon_temperature}.

\section{Summary}
\label{sec:summary}

To summarize, we have studied the heat balance in a weakly disordered system with local inhomogeneities. Specifically,  we have considered a spherical local constriction with increased impurity scattering rate as compared to the scattering in the uniform background.
We assumed that electron-electron interaction is strong in two senses: firstly, fast electron-electron collisions drive the system into the hydrodynamic regime; secondly, it  guarantees the electrical neutrality, thus  making the electron liquid nearly incommpressible.

In the absence of viscosity,  the electron temperature  distribution induced  by the inhomogeneity is described by a Landauer-dipole-like structure that is shifted along the current by the amount of the order of the inelastic drift length (Fig.~\ref{fig:large_ec}).
The Landauer-dipole distribution stems from the tendency of particles to travel around  the inhomogeneity (in the 2D and 3D cases).
Remarkably, the thermal Landauer dipole and its asymmetry, both induced by the current flow,  exist in systems of arbitrary dimensionality, and of genuine constriction geometries. Thus, we have found that
the heating of inhomogeneous ideal electron fluid is universally described by a Landauer dipole, which is deformed by the flow beyond the linear-response regime.

Further, focusing on the 2D case, we have explored the evolution of the electron temperature and velocity profiles with increasing viscosity of the electron fluid (Fig.~\ref{fig:vis_temp_2d}). Our main conclusion is that the viscosity dramatically changes the heat balance in the system. Specifically, we have found  that viscosity suppresses the Landauer-dipole structure and---for sufficiently high viscosity---essentially reduces the heat-balance problem in all dimensions to a quasi-1D problem. Most importantly, viscosity leads to the formation of the ``hot spot'' exactly in the position of the  inhomogeneity. We have also derived a relation between  electron temperature and the temperature of the phonon system (Fig.~\ref{fig:phonon_temperature}), which can be directly measured in experiment.

\acknowledgments
\label{Ack}
We thank A. Dmitriev, A. Mirlin, B. Narozhny, and E. Zeldov for discussions.
The work of GZ and IG is supported by the German-Israeli Foundation (GIF grant No.~I-1505-303.10/2019), the DFG grant (No.~GO 1405/5) within the FLAG-ERA Joint Transnational Call (project GRANSPORT), and European Commission under the EU Horizon 2020 MSCA RISE-2019 program (project 873028 HYDROTRONICS). The work of VK and IG was supported by the Russian Science Foundation (Grant No.~20-12-00147) . The work of VK was also supported by the Foundation for the Advancement of Theoretical Physics and Mathematics ``BASIS''.

\appendix
\section{Gaussian constriction} \label{Gauss}

In this Appendix, we derive the expressions for Landauer-dipole structures in $D=1,2,3$ dimensions. To this end, one needs to calculate the Fourier transform of $K_\m q \xi_\m q$,  where $K_\m q$ and $\xi_\m q$ are given by Eqs.~\eqref{Kq} and \eqref{eq:circular_constriction} of the main text, respectively.  
Introducing auxiliary integrals,
$$
\frac{1}{q^2}= \int \limits_0^\infty dt\, 
e^{- t q^2},\quad 
\frac{1}{i q_\parallel +1/l_{\rm in}} =  
\int \limits_0^\infty ds 
e^{- s(i q_\parallel+ 1/l_{\rm in})},
$$
we find 
\be
\begin{aligned}
& \int\!\! \frac{ d^D \m q}{(2 \pi)^D}  \frac{q_\parallel^2-q_\perp^2}
{q_\parallel^2+q_\perp^2} 
\frac{e^{i \m q \m r -q^2L^2/4}}{i q_\parallel+1/l_{\rm in}}
\\
&=\!
\int \limits_0^\infty\! ds  e^{- s /l_{\rm in}} \!\! 
\int \limits_0^\infty\! dt (\p_\perp^2-\p_\parallel^2) 
\!\int\!\! \frac{ d^D \m q}{(2 \pi)^D} e^{-q^2 (t+L^2/4) + i \m q \m r - i q_\parallel  s }
\\
&
=\!\int \limits_0^\infty ds e^{- s /l_{\rm in}} \tilde Q_D (x - s, \m{r}_{\perp}),
\end{aligned}
\ee
where 
\be 
\tilde Q_D (\m{r}) =\frac{1}{ \pi ^{D/2}}  (\p_\perp^2-\p_\parallel^2) \int \limits_0^\infty dt \frac{ e^{-r^2/(4 t + L^2)}}{ 4 t + L^2  } .
\ee
Applying the spatial derivatives $(\p_\perp^2-\p_\parallel^2)$, 
we get 
\begin{align}
\tilde Q_D (\m{r}) &= \int \limits_ {L^2/4}^\infty dt e^{- r^2/4t} 
\notag
\\
&\times 
\left\{ 
\begin{array}{cc}
      \dfrac{2 t-x_\parallel^2}{8 \sqrt{\pi} t^{5/2}}  ,  \quad &\text{for} \quad D=1 , \\[0.3cm]     \dfrac{r_\perp^2-x_\parallel^2}{16  \pi  t^{3}}  ,  \quad &\text{for} \quad D=2 ,
     \\[0.3cm]  
      \dfrac{r_\perp^2-x_\parallel^2- 2 t}{32 \pi^{3/2} t^{7/2}} ,  \quad & \text{for} \quad D=3. 
      \end{array} 
      \right.
\end{align}
Calculating the remaining integral over $t$ and assuming that electric field is parallel to the $x$-axis, we arrive at  Eq.~\eqref{tildeQ} of the main text.

\section{Temperature Landauer dipole in 3D and 2D system---a local heating approximation}
\label{app:multi_conduct}

\begin{figure}
  \centering
      \includegraphics[width=0.9\columnwidth]{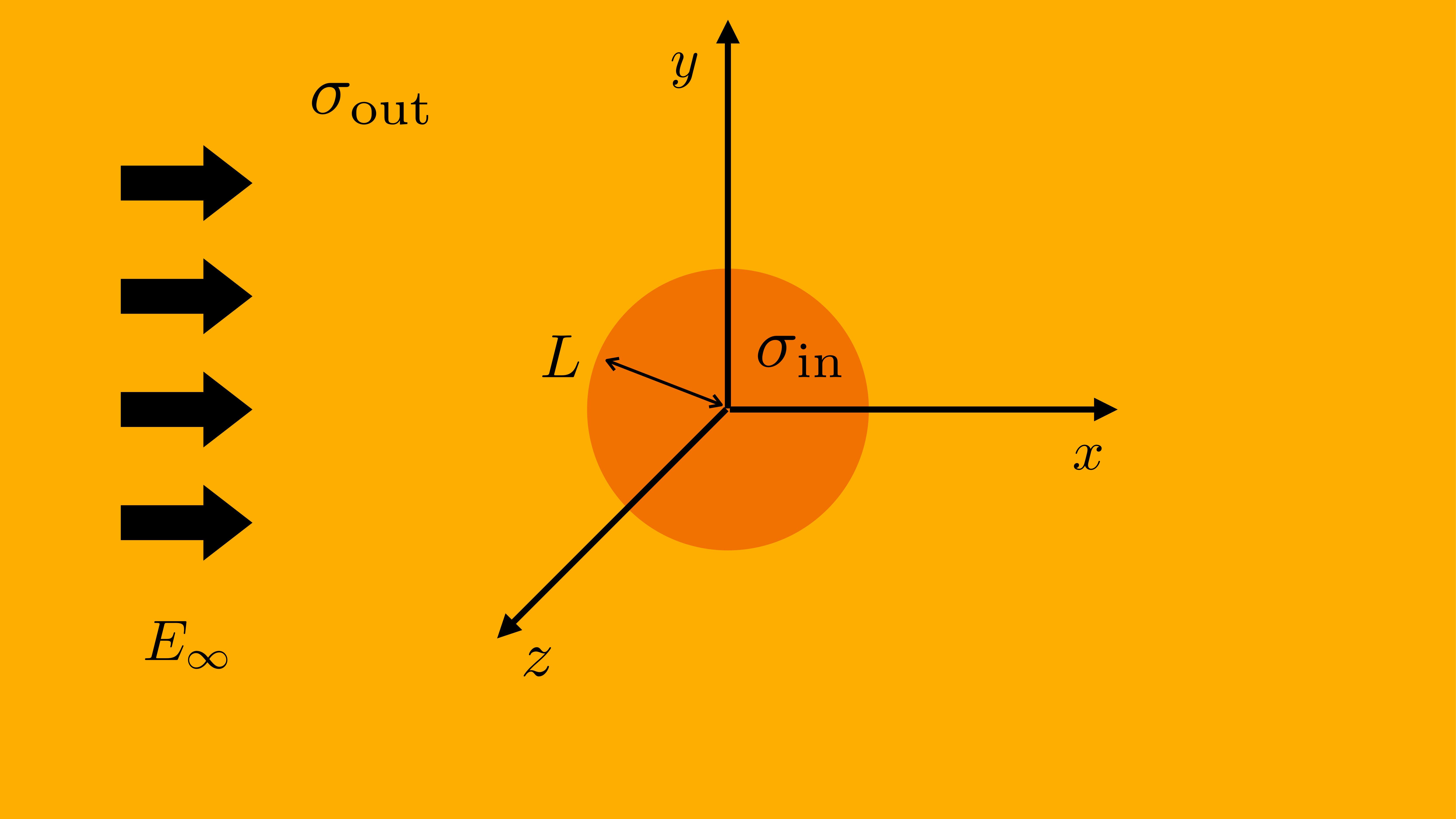}
  \caption{Phenomenological local-conductivity model adopted for the local Joule approximation. A constriction of radius $L$, located at the origin, is characterized by the conductivity $\sigma_{\text{in}}$ that is smaller than the uniform background conductivity: $\sigma_{\text{in}} < \sigma_{\text{out}}$. The external field $E_{\infty}$ is directed along the $x$-axis.}
  \label{fig:3d}
\end{figure}

In this Appendix, we employ the local heating approximation (Fig.\,\ref{fig:3d}), and study the constriction-induced local heating of 2D and 3D systems. The calculation here generalizes the one in Ref.~\cite{TikhononvPRB18}.

\subsection{Landauer dipole in 3D}
\label{app:3d}

Within the local heating approximation, the constriction of radius $L$ has the uniform conductivity $\sigma_{\text{in}}$ that is smaller than the conductivity $\sigma_{\text{out}}$ outside the constricted area. The electrical current is driven by the field $E_{\infty}$ directed along the $x$-axis.
Under the continuity boundary condition, the electric potential becomes
\begin{equation}
\begin{aligned}
\varphi(\m r)
= \left\{
\begin{array}{ll} 
\displaystyle  
-\frac{3 \sigma_{\text{out}}}{\sigma_{\text{in}} + 2 \sigma_{\text{out}}}\, r E_{\infty}  \cos\vartheta,  & r < L, 
\\[0.5cm]  
\displaystyle - 
\left[ r +\frac{(\sigma_{\text{out}} - \sigma_{\text{in}}  ) L^3}{(2\sigma_{\text{out}} + \sigma_{\text{in}} ) r^2}\right] 
E_{\infty}  \cos\vartheta,\quad & r > L, \end{array}\right. 
\end{aligned}
\label{eq:efield}
\end{equation}
where $\vartheta$ is the angle between the $x$ axis and the direction of the vector $\m r$ in real space.
Apparently, $\varphi(\m r)$ reduces to the uniform value $- E_{\infty} r \cos\vartheta$ in the $r\equiv |\m r| \gg L$ limit.

With the electric potential from Eq.~\eqref{eq:efield}, we find the local electric field and calculate the local Joule heating $\sigma(\m r) E(\m r)^2$
\begin{equation}
\begin{aligned}
& \frac{P(\m r)}{\sigma_{\text{out}} E_{\infty}^2} = \frac{\sigma(\m r) E(\m r)^2}{\sigma_{\text{out}} E_{\infty}^2}\\
& = \left\{\begin{array}{ll} 
\displaystyle (1 + \lambda - 2\lambda^2)
\sigma_{\text{out}} E_{\infty}^2,  & r < L, \\ \\  \displaystyle \!\!\left[ 1 \!-\! \frac{L^3}{r^3} \lambda (1 \!- \!3 \cos\vartheta) \right]^2\!\! +\! \frac{9L^6}{4r^6} \lambda^2 \sin^2(2\vartheta),\ & r > L, \end{array}\right. 
\end{aligned}
\label{eq:local_heating}
\end{equation}
where
\begin{equation}
\lambda \equiv \frac{\sigma_{\text{in}} - \sigma_{\text{out}}}{2 \sigma_{\text{out}} + \sigma_{\text{in}}}
\end{equation}
is the parameter characterizing the inhomogeneity. It vanishes in the homogeneous situation 
($\sigma_{\text{in}}=\sigma_{\text{out}}$), and approaches $1/2$ when the constriction is insulating ($\sigma_{\text{in}} \to 0$).

To describe the effect of the constriction, we define 
$$\delta P = P - \sigma_{\text{out}} E_{\infty}^2,$$ 
as the constriction-induced extra Joule heating
and express it in terms of $\lambda$:
\begin{equation}
\begin{aligned}
&\frac{\delta P}{\sigma_{\text{out}} E_{\infty}^2}\\
& \!=\! \left\{\begin{array}{ll} \displaystyle \!\lambda - 2 \lambda^2,  & r < L, \\ \\  \displaystyle  \! \frac{\lambda L^3 [1 + 3 \cos(2 \vartheta)] }{r^3} 
\! + \!\frac{\lambda^2 L^6[5\! + \!3 \cos(2\vartheta)]}{2 r^6}\!,\ & r > L. \end{array}\right. 
\end{aligned}
\label{eq:local_heating_r}
\end{equation}
Clearly, $\delta P$ vanishes when $\lambda = 0$. It is also non-monotonous and approaches the peak value when $\lambda = 1/4$. Indeed, all particles detour around the constriction in the insulating limit, where the Joule heating vanishes.

For a more intuitive understanding, we provide the profile of the local Joule heating (when $\lambda = 1/3$) in Fig.~\ref{fig:model}(d) of the main text. Apparently, a Landauer-dipole heating pattern emerges around the constriction. This heating pattern naturally induces a shifted Landauer dipole of the electron temperature profile. Noteworthy, this Landauer-dipole feature is produced by the residual charges at the boundary of the constriction. These charges also guarantee the constant electric field in the constricted area.

\subsection{Landauer dipole in 2D }
\label{app:2d_multi_conductance}

Let us move to the 2D case with a circular-shape constriction. To begin with, following the same method as that of Appendix\,\ref{app:3d}, we arrive at the electric field of 2D
\begin{equation}
\begin{aligned}
\m E(\m r) 
& = \left\{\begin{array}{ll} \displaystyle  \m E_{\infty}\! + \delta \m E_\text{in},  \ & r < L, \\ \\  \displaystyle \m E_{\infty}\! +\! \delta \m E_{\text{out}}, \ & r > L, \end{array}\right. 
\end{aligned}
\label{eq:2d_e_field}
\end{equation}
where
\begin{equation}
    (\delta \m E_\text{in})_x = \delta E,\ \ \ \ (\delta\m E_\text{in})_y = 0
  \label{eq:inhomogenous_field_in}
\end{equation}
is the electric field in the constricted area, and
\begin{equation}
    \begin{aligned}
        (\delta \m E_\text{out})_x & = \delta E \frac{y^2 - x^2}{(x^2 + y^2)^2}L^2,\\
        (\delta \m E_\text{out})_y & = \delta E \frac{-2xy}{(x^2 + y^2)^2}L^2,
     \end{aligned}
     \label{eq:inhomogenous_field_out}
\end{equation}
refers to that outside of the constriction.
In Eqs.~\eqref{eq:inhomogenous_field_in} and \eqref{eq:inhomogenous_field_out}, the inhomogeneity-induced electric field is given by
\begin{equation}
    \delta E = E_{\infty} \frac{\sigma_{\text{out}} - \sigma_{\text{in}}}{\sigma_{\text{out}} + \sigma_{\text{in}}} = E_{\infty} \lambda_2 >0,
    \label{eq:de}
\end{equation}
where the parameter $\lambda_2$ characterizes the strength of the inhomogeneity in the 2D case.

\begin{figure}
  \centering
      \includegraphics[width=0.45 \textwidth]{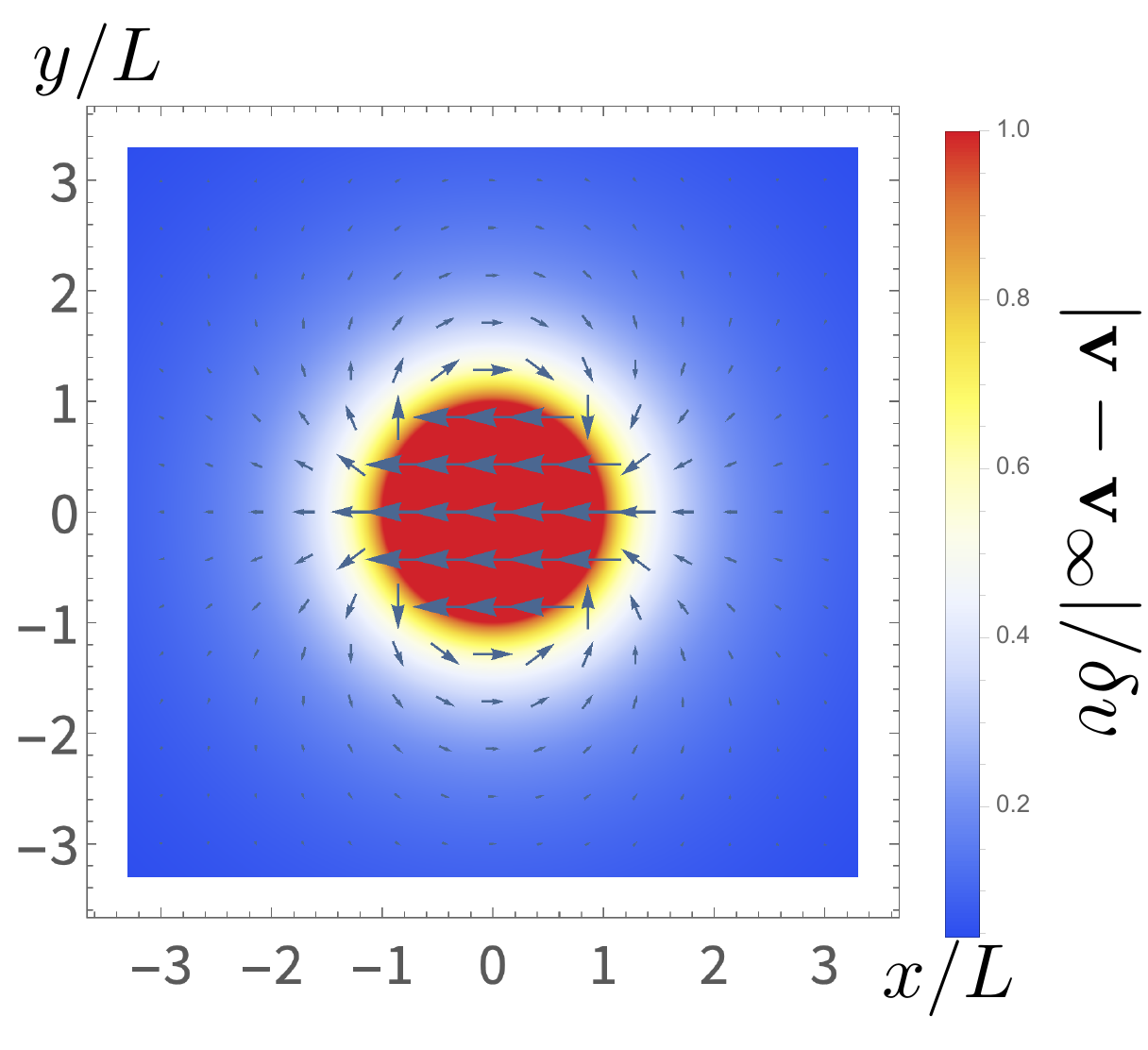}
  \caption{The vector velocity profile of the 2D local heating model.}
  \label{fig:2d_velocity_phenom}
\end{figure}

In the local approximation, the drift velocity is proportional to the electric field 
$\m v(\m r) = \sigma(\m r) \m E(\m r)/N_\infty e$. Following Eq.~\eqref{eq:2d_e_field}, we obtain the drift velocity profile
\begin{equation}
\begin{aligned}
\m v(\m r)
& = \left\{\begin{array}{ll} \displaystyle  \m v_{\infty} +\delta \m v_\text{in},  & \ r < L, \\ \\  \displaystyle \m v_{\infty} + \delta \m v_\text{out} , & \ r > L, \end{array}\right. 
\end{aligned}
\label{eq:e_velocity}
\end{equation}
where
\begin{equation}
    (\delta \m v_\text{in})_x = -\delta v,\ \ (\delta \m v_\text{in})_y = 0
    \label{eq:v_in}
\end{equation}
in the constricted area, and
\begin{equation}
    \begin{aligned}
      & (\delta \m v_\text{out})_x = \delta v \frac{y^2-x^2}{(x^2 + y^2)^2} L^2\\
      & (\delta \m v_\text{out})_y = \delta v \frac{-2 xy}{(x^2 + y^2)^2} L^2
    \end{aligned}
    \label{eq:v_out}
\end{equation}
outside of the constriction.
In Eqs.~\eqref{eq:v_out} and \eqref{eq:v_in}, the homogeneous velocity $\m v_{\infty} = \sigma_{\text{out}}\m E_\infty/N e$, and
\begin{equation}
\delta v = v_{\infty} \frac{\sigma_{\text{out}} - \sigma_{\text{in}}}{\sigma_{\text{out}} + \sigma_{\text{in}}} \equiv v_{\infty} \lambda_2
\label{eq:dv}
\end{equation}
is the inhomogeneity-induced velocity variation, defined in the same manner as Eq.~\eqref{eq:de}.
In the momentum space, the variation of velocity becomes
\begin{equation}
\begin{aligned}
\delta v_x (\m q) & = \frac{2\pi L}{q} [  J_1(L q) \cos (2\varphi)  - J_1(L q) ]  \lambda_2 \\
\delta v_y (\m q) 
& = \frac{2\pi L}{q}  J_1(L q) \sin (2\varphi)   \lambda_2,
\end{aligned}
\label{eq:dv_momentum}
\end{equation}
where $q = |\m q|$, the angle $\varphi$ refers to the momentum direction, and $J_1(Lq)$ is the Bessel function.

To better understand Eq.~\eqref{eq:dv_momentum}, we Fourier transform  the circular-shape step function, $\xi (\m r) = \xi_0 \Theta (x^2 + y^2 - L^2)$, into the expression in the momentum space
\begin{equation}
    \xi(\m q) = \xi_0 \frac{2 \pi}{q} L J_1(L q).
    \label{eq:ft_step_function}
\end{equation}
The combination of Eqs.~\eqref{eq:dv_momentum} and \eqref{eq:ft_step_function} gives us the velocity kernel of the local-heating model
\begin{equation}
    \begin{aligned}
    K_{u_x} & = \frac{-2 q_{\perp}^2}{q_\parallel^2 + q_\perp^2},\\
    K_{u_y} & = \frac{2 q_\parallel q_\perp}{q_\parallel^2 + q_\perp^2},
    \end{aligned}
    \label{eq:kernel_local_heating_model}
\end{equation}
where $q_{\parallel}$ and $q_\perp$ refer to two momentum components.
Equation~\eqref{eq:kernel_local_heating_model} coincides with Eq.~\eqref{dv-final} of the main text, except for an overall extra factor of two.
This extra factor is related to the definition of $\lambda_2$: at the weak inhomogeneity limit, $\lambda_2 \approx (\sigma_{\text{out}} - \sigma_{\text{in}})/2\sigma_{\text{out}}$.
Following Eq.~\eqref{eq:dv}, we plot the 2D velocity profile in Fig.~\ref{fig:2d_velocity_phenom}.
It captures the major feature of that in the main text: particles tend to detour around the constriction.

\begin{figure}[h!]
  \centering
      \includegraphics[width= 1 \columnwidth]{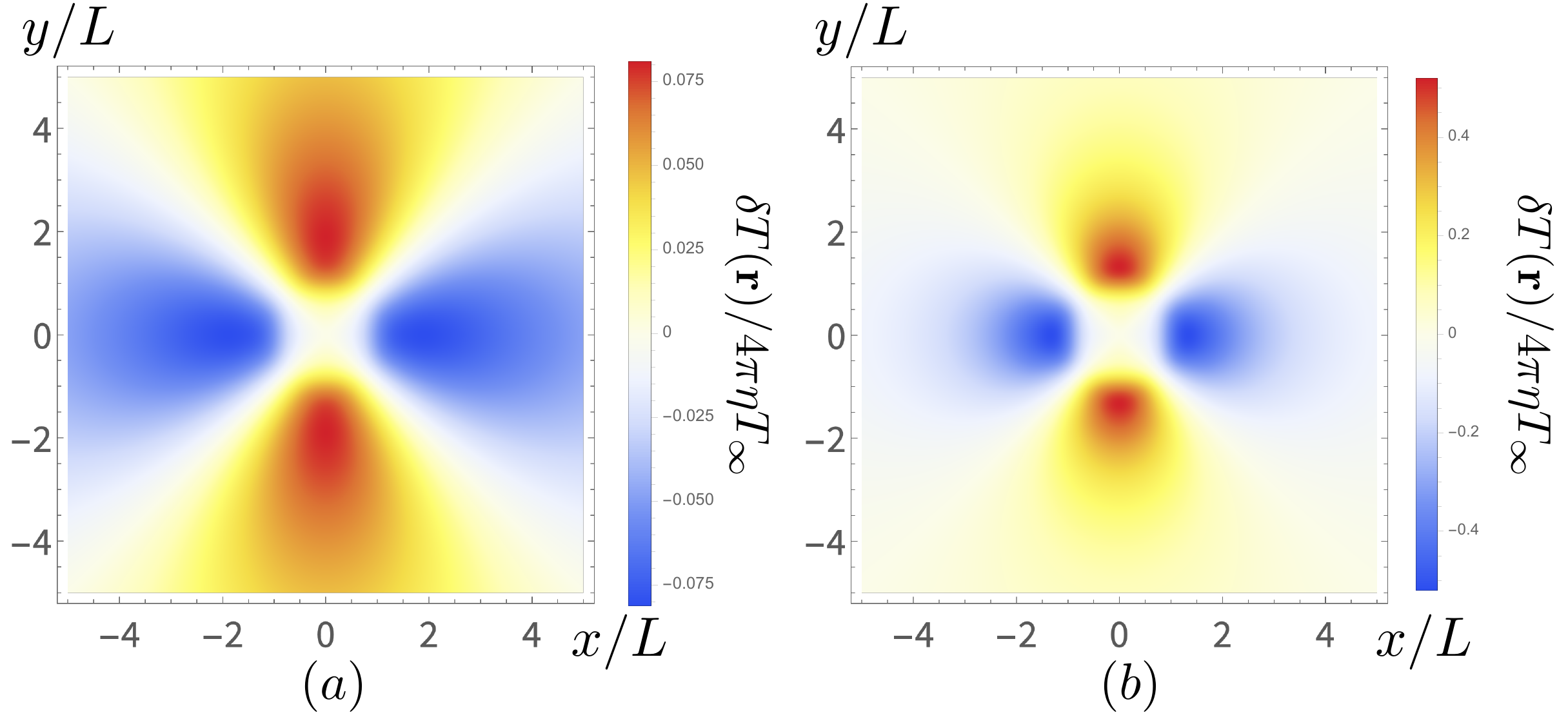}
  \caption{The temperature profiles of the phenomenological model when (a) $l_* = 2 L$; (b) $l_* = 0.5 L$. The temperature profiles of the diffusive phenomenological model are completely symmetric.}
  \label{fig:diffusive_t}
\end{figure}

With the velocity profile known, we can calculate the temperature profile.
Since this local-heating model is diffusive, $\varkappa_{\text{e}} \gg v_{\infty}$ (where $\varkappa_{\text{e}} $ is the electronic heat conductivity), we keep only the diffusive term in the heat diffusion equation
\begin{equation}
-\varkappa_{e} \bigtriangledown^2 \delta T = ne \m E \m v - \gamma \delta T.
\end{equation}
As a simple check, we focus on the weak-inhomogeneity limit $\lambda_2 \ll 1$.
In this limit, the Fourier-transformed temperature profile becomes
\begin{equation}
\begin{aligned}
  \delta T_{\m q} &= T_\infty  \frac{4\pi}{q} \frac{ L J_1(Lq) \cos (2\varphi) }{ l_*^2 q^2 + 1} \lambda_2,
  \end{aligned}
   \label{eq:2d_temp_phenom}
\end{equation}
where $l_* = \sqrt{\varkappa_{\text{e}} /\gamma}$.
In Eq.~\eqref{eq:2d_temp_phenom}, the prefactor $T_\infty$ is the homogeneous temperature when $\lambda_2 = 0$.
We present the temperature profile of Eq.~\eqref{eq:2d_temp_phenom} in the real space, in Fig.~\ref{fig:diffusive_t}.
In contrast to the hydrodynamic model of the main text Eq.~\eqref{dT-final}, the temperature kernel of Eq.~\eqref{eq:2d_temp_phenom} strictly corresponds to a Landauer dipole that sits exactly at the center of the constricted area: the asymmetry thus completely disappears in the diffusive situation.

\bibliography{References_2D.bib}

\end{document}